\documentclass[prc,aps,preprint,nofootinbib,showkeys,eqsecnum,tightenlines]{revtex4}    

\usepackage{graphicx}
\usepackage{dcolumn}

\usepackage{bm}

\usepackage{latexsym,amsmath,amssymb}
\usepackage{amssymb}
\usepackage{amsmath}
\usepackage{amsfonts}
\usepackage{verbatim}
\usepackage{slashed}

\def\be{\begin{equation}}
\def\ee{\end{equation}}
\def\bea{\begin{eqnarray}}
\def\eea{\end{eqnarray}}

\def\vk{\hat{\bm k}}
\def\vkp{\hat{\bm k}'}
\def\vp{\bm p_f }
\def\vpp{\bm p_{\mathrm{e}^+}}
\newcommand{\noi}{\noindent}

\begin{document}
%
%
\title{Consequence of total lepton number violation in strongly magnetized iron white dwarfs}
\author{V.B.~Belyaev\,\footnote{deceased}}\affiliation{Bogolyubov Laboratory of Theoretical Physics, Joint
Institute for Nuclear Research, Dubna  141980, Russia}
\author{P. Ricci}
\affiliation{Istituto Nazionale di Fisica Nucleare, Sezione di
Firenze, I-50019 Sesto Fiorentino (Firenze), Italy }
\author{F.~\v{S}imkovic}\affiliation{Department of Nuclear Physics and Biophysics,
Comenius University, Mlynsk\'a dolina F1, SK--842 15, Bratislava,
Slovakia and  Bogolyubov Laboratory of Theoretical Physics, Joint
Institute for Nuclear Research, Dubna  141980, Russia}
\author{J.~Adam~Jr.}
\affiliation{Institute of Nuclear Physics ASCR, CZ--250 68
\v{R}e\v{z}, Czech Republic }
\author{M.~Tater}
\affiliation{Institute of Nuclear Physics ASCR, CZ--250 68
\v{R}e\v{z}, Czech Republic }
\author{E.~Truhl\'{\i}k\,\footnote{Corresponding author}\,\footnote{E-mail address: truhlik@ujf.cas.cz (E. Truhl\'ik)}}
\affiliation{Institute of Nuclear Physics ASCR, CZ--250 68
\v{R}e\v{z}, Czech Republic}

\begin{abstract}
The influence of a neutrinoless electron to positron conversion on a
cooling of strongly magnetized iron white dwarfs is studied. It is
shown that they can be good candidates for soft gamma-ray repeaters
and anomalous X-ray pulsars.
\end{abstract}


\noi \hskip 1.9cm 

\maketitle

\section{Introduction}
\label{intro}

The white dwarfs (WDs) are  quite numerous in the Milky Way
\cite{YT} and the astrophysics of these compact objects is nowadays
a well developed domain \cite{KC,BH,ACIG}. Since the interior of the
WDs is considered to be fully degenerate, studying their properties
provides a fundamental test of the concept of stellar degeneracy.
The important observables, which could test models of structure and
evolution of WDs, are the luminosity and the effective (surface)
temperature.

Steady progress in understanding of the WDs cooling processes and
precise measurements of their luminosity curve and of their
effective temperature open a door to their possible use as a
laboratory for analyzing some problems of elementary particles
physics. Thus, Isern et al. \cite{IEA1,IEA2} suggested studying
possible existence of axions on a basis of the WDs luminosity
function. Following this idea, we analyze the influence of the
lepton number violation on the luminosity and the effective
temperature of strongly magnetized iron WDs (SMIWDs)\footnote{We
will use acronyms WDs (SMWDs) for the white dwarfs with the magnetic
field of ${\cal B}\ll{\cal B}_c$=4.414 $\times 10^{13}$ G (${\cal
B}\ge{\cal B}_c$), respectively.}. As it is well known, the
existence of the Majorana type neutrino would imply the lepton
number violating process of electron capture by a nucleus X($A,Z$)
\begin{equation}
\mathrm{e}^- + \mathrm{X}(A,Z) \rightarrow \mathrm{X}(A,Z-2) +
\mathrm{e}^+ \ . \label{EMEPC}
\end{equation}
This reaction is an analogue of the neutrinoless double beta-decay,
intensively studied  these days \cite{VES}. Very recent experimental
results on this process, obtained with the $^{76}$Ge detectors, can
be found in Refs.\,\cite{GD,MAJ}. An estimate \cite{MAJ} shows that
to attain for the neutrino masses sensitivities in the region of
15\,-\,50 meV, tonne-scale detectors are needed. At present the
detectors \cite{GD,MAJ} comprise tens of kilograms of $^{76}$Ge.
Since 1 kg of $^{76}$Ge includes 8.31\,$\times$\,10$^{24}$ atoms, a
tonne device would contain $\sim$ 10$^{28}$ of $^{76}$Ge atoms. On
the other hand, the matter density of the SMIWDs is at the level of
10$^{33}$/cm$^3$ and more, which is by several orders of magnitude
larger. This fact makes the study of reaction (\ref{EMEPC}) in
stellar medium attractive.

For the weak reaction (\ref{EMEPC}) with the rate proportional to
the square of a small neutrino mass to be detectable, it has to take
place in a bulk of the stellar body with a well understood
background. It is allowed energetically when the Fermi energy
$E_{\,\mathrm{F}}$ of the electron gas is larger than the threshold
energy $\Delta^{\beta\beta}_Z$ given by the mass difference between
the final and the initial nuclei plus the electron mass.  However,
as it can be seen from Table \ref{tab:pon}, in the WDs consisting of
the even-even nuclei the threshold energy of the inverse beta decay
$\Delta^\beta_Z$ is smaller than $\Delta^{\beta\beta}_Z$ +
$m_\mathrm{e}c^2$ ($m_\mathrm{e}$ is the electron mass and $c$ is
the light velocity). Since usually
$\Delta^\beta_{Z-1}\,<\,\Delta^\beta_Z$, two successive decays
\cite{EES,ST,RRRX,BK1,BS}
\begin{equation}
(A,Z)\,\rightarrow\,(A,Z-1)\,\rightarrow\,(A,Z-2)\, \label{IBD}
\end{equation}
proceed, unless all $(A,Z)$ nuclei transform to $(A,Z-2)$ nuclei,
and the reaction (\ref{EMEPC}) cannot occur. The point is that in
the WDs, the electron Fermi energy $E_{\,\mathrm{F}}$ cannot
overcome the energy at which the inverse beta decay proceeds. For
the case of the $^{56}_{26}$Fe nuclei, this situation is discussed
in detail in Ch.\,3 of Ref.\,\cite{ST}. Instead of compression
increasing $E_{\,\mathrm{F}}$ and, therefore, the pressure, the
electrons are captured by the iron nuclei which are transformed in
the two-step process (\ref{IBD}) into the chrome ones. So the onset
of the inverse beta decay at the density $\rho_B$=(1.14 - 1.18)
$\times$ 10$^9$ g/cm$^{3}$ terminates the iron WD \cite{ST,RRRX}. As
can be seen from the Table III of Ref.\,\cite{RRRX}, this density is
also critical one for the onset of the instability due to the
general relativity and the subsequent collapse of the $^{56}_{24}$Cr
WD happens.\footnote{ See also the discussion in the paragraph
containing Eq.\,(\ref{R1}) below.} So, the only chance to have the
reaction (\ref{EMEPC}) in the bulk of compact objects are the SMWDs,
where it can hold \mbox{$E_{\,\mathrm{F}}$ $>$
$\Delta^{\beta\beta}_Z$ + $m_\mathrm{e}c^2$} due to the presence of
the strong magnetic field.

Besides our choice for X($A,Z$)\,=\,$^{56}_{26}$Fe(0,0$^+$) and for
X($A,Z-2$)\,=\,$^{56}_{24}$Cr(0,0$^+$) in the process (\ref{EMEPC}),
other pairs of the even-even nuclei in the ground state (0,0$^+$)
can be taken. They are given in Table \ref{tab:pon}, together with
the value of the threshold energies $\Delta^{\beta\beta}_Z$ and
$\Delta^\beta_Z$. Since the energy $E_e$ of electrons that can
participate in the reaction (\ref{EMEPC}) should satisfy inequality
\be
m_\mathrm{e}c^2 + \Delta^{\beta\beta}_Z < E_\mathrm{e} <
E_{\,\mathrm{F}}\,, \label{INQ}
\ee
it follows from Table \ref{tab:pon} that there are more active
electrons for the nuclei with smaller values of
$\Delta^{\beta\beta}_Z $. On the other hand, the inverse beta decay
process cannot occur in the nucleus $^{12}_{\,\,\,6}$C in reality.
This follows from comparing the critical densities for the onset of
the inverse beta decay
$\rho^{\beta,\mathrm{relFMT}}_{\mathrm{crit}}$ and for the
gravitational instability $\rho^{\mathrm{FMTrel}}_{\mathrm{crit}}$ ,
as given in Table II and Table III \cite{RRRX}, respectively,
\be
\rho^{\beta,\mathrm{relFMT}}_{\mathrm{crit}} =
3.97\,\times\,10^{10}\,\mathrm{g/cm^3}
\,,\,\,\,\rho^{\mathrm{FMTrel}}_{\mathrm{crit}}
=2.12\,\times\,10^{10}\,\mathrm{g/cm^3}\,.   \label{RCRIT}
\ee
It also follows from the inequality (\ref{INQ}) and from Table
\ref{tab:pon} that the reaction (\ref{EMEPC}) would be strongly
suppressed in the $^{16}_{\,\,\,8}$O SMWDs in comparison with the
case of SMWDs composed of heavier nuclei.

Let us also note that our choice of the SMIWDs was influenced by the
existence of extensive calculations of the properties of the iron
WDs for the matter densities small enough to avoid the inverse beta
decay \cite{PAB} and also for the case when the inverse beta decay
takes place \cite{BK1,BS,BK2}. This will allow us to make a
qualitative comparison of our results with these already existing
ones.

The non-magnetized iron-core WDs were first studied in
Ref.~\cite{SHV}. A comprehensive study of the properties of
iron-core WDs with emphasis placed on their evolution was performed
in Ref.\,\cite{PAB}. In particular, the crystallization,
electrostatic corrections to the equation of state, conductive
opacity and neutrino emission were taken into account. The work
\cite{PAB} was inspired by new observational data provided by the
satellite Hypparcos, from which the existence of three iron-core WDs
(GD 140, EG 50 and Procyon B) was suggested \cite{PROV1,HLSJLP}. For
Procyon B, being situated close to Procyon A, one of the brightest
stars in the sky, it was difficult to obtain good data and later on,
it was put 'outside the iron box' and classified as a rare DQZ WD
\cite{PROV2}.  For EG 50 and GD 140 the results were robust enough
for Provencal et al. to conclude  that the only way how to explain
the observations was to assume an iron, or an iron-rich, core
composition of these two WDs. Only then it was possible to fit the
observed radii, masses, and surface gravities consistently. This
opened the problem of understanding formation of these WDs, because
such a core composition is at variance  with current theories.

Later on, Fontaine et al. \cite{FBB} reconsidered the problem with
the EG 50 core composition by improving the accuracy of the
effective temperature $T_{\mathrm{eff}}$ and surface gravity $g$ of
the EG 50 deduced from optical spectroscopy. Since these values
turned out to be entirely consistent with those used in
Ref.\,\cite{PROV1}, they concluded that the problem with the
iron-core of EG 50 does not lie in inaccuracy of spectroscopic data.
Subsequent calculations \cite{FBB} of the distance with various core
compositions and its comparison with the observed distances $d =
 15.41  \begin{array}{c} +0.84 \\ -0.71 \end{array}$ pc, provided by
Hypparcos \cite{PROV1}, and $d = 15.08  \begin{array}{c} {+0.50} \\
{-0.40} \end{array}$ pc from the Yale Parallax Program, show (see
Fig.\,2 \cite{FBB}) that cores made of C, O, Ne, Mg, Si, S and Ca
are excluded and only models with heavier element cores Ar, Ti, Cr,
Fe provide acceptable solutions. Then Fontaine et al. conclude that
the case of EG 50 shows that the WD formation process is not fully
understood.

It can be concluded from the discussed material that the main source
of uncertainty of the core composition of EG 50 is in the parallax.
It is expected that the uncertainty will diminish essentially with
the new precise data from GAIA \cite{MABEAL}.

It also follows that our choice of the SMIWDs is well founded,
because calculations comparable with Refs.\, \cite{PAB,BK1,BS,BK2}
for other possible elements do not exist.

Nowadays, two possible physical processes able to account for the
formation of iron core WDs are available \cite{ICL,OSJ}. In
Ref.\,\cite{OSJ}, namely the case of the WD EG 50 is considered and
a simple model for the explanation of its Fe-rich composition is
proposed. If a low-mass X-ray binary, consisting of a neutron star
and a WD, is sufficiently tight, under certain assumptions, ejecta
from the exploding neutron star trigger nuclear burning in the WD,
possibly leading to the WD with Fe-rich composition with the mass
$0.43 M_\odot < M_{\mathrm{WD}} < 0.72 M_\odot$\footnote{Here,
$M_\odot \approx$ 2$\times$ 10$^{33}$ g is the solar mass.},
reminiscent of EG 50.

Rather exotic solution of the above discussed problem has been
proposed in Ref.\,\cite{MEAL}, considering possibility that such
more compact WDs as ED 50 or GD 140 are (characterized by falling
away from the expected C/O relationship in the M-R diagram), could
have in the core a portion of strange matter gained after accreting
a strange-matter nugget. Such nuggets could exist (see
Ref.\,\cite{MEAL} and the references therein) either as a relic of
the early universe or as an ejected fragment from the
merger/coalescence of strange-matter neutron stars. However, as is
well known \cite{ST,PPNS}, the strange matter could be present only
in the inner core of rather massive NSs with the mass $M \gtrsim$
1.4 - 1.5 $M_\odot$, where the matter density $\rho\,\gtrsim
2\,\rho_0$\footnote{The normal nuclear density $\rho_0$ = 2.8
$\times$ 10$^{14}$ g/cm$^{3}$.}. So it is not clear how could be the
strange matter maintained in the core of WDs that have the central
density smaller by several orders in the magnitude.

Let us also present the study \cite{CRIG} in which a common proper
motion pair formed by a WD0433+270 and a main-sequence star BD+26730
is studied with aim to investigate whether this system belongs to
the Hyades cluster. In the affirmative case, the calculations of
cooling sequence for different core compositions based on the
results of \cite{PAB} show that the WD member of the pair could have
an iron core. The kinematic and chemical composition considerations
provided enough material for Catalan et al. to make believe that the
pair was a former member of the Hyades cluster and consequently, it
has an evolutionary link to it. However, the evidence is not yet
fully conclusive.

\begin{table}[tb]
\caption{ The pairs of the even-even nuclei with which the reaction
(\ref{EMEPC}) can proceed, for which $\Delta^{\beta\beta}_Z$ are the
threshold energies, calculated using the atomic mass evaluation
\cite{AWT} ; $\Delta^\beta_Z$ are the threshold energies of the
inverse beta decay. Their values were taken from
Refs.\,\cite{BK1,BS}. In the 5th, 8th and 9th rows, they contain
also the excitation energies of the 1$^+$ level of the related
$(A,Z-1)$ nuclei, because the transitions to the ground level are
strongly suppressed.
 }
\begin{center}
\begin{tabular}{| c  |  c    | c    |  c  | }\hline
       X($A,Z$) &  X($A,Z-2$)   & $\Delta^{\beta\beta}_Z$ [MeV] & $\Delta^\beta_Z$ [MeV] \\\hline
 $^{12}_{\,\,\,6}$C     & $^{12}_{\,\,\,4}$Be   & 27.0  & 13.370 \\
 $^{16}_{\,\,\,8}$O     & $^{16}_{\,\,\,6}$C    & 19.45 & 10.419 \\
 $^{32}_{16}$S          & $^{32}_{14}$Si  & 2.96  &  1.708 \\
 $^{56}_{26}$Fe         & $^{56}_{24}$Cr  & 6.33  &  3.794 \\
 $^{42}_{20}$Ca         & $^{42}_{18}$Ar  & 5.15  &  3.524 \\
 $^{60}_{28}$Ni         & $^{60}_{26}$Fe  & 4.08  &  2.890 \\
 $^{66}_{30}$Zn         & $^{66}_{28}$Ni  & 3.92  &  2.630 \\
 $^{72}_{32}$Ge         & $^{72}_{30}$Zn  & 5.48  &  4.260 \\\hline
\end{tabular}
\end{center}
\label{tab:pon}
\end{table}

While considering the process (\ref{EMEPC}) in SMIWDs is
theoretically tempting, it should be pointed out that their
astrophysical studies have only started to develop. In particular,
microscopic models of their development, of their internal structure
or of their cooling process have not yet been developed to a point
of general acceptance and to convincing tests by astrophysical data,
though the very recent publications \cite{BEBH,DM6} have made a
basic breakthrough in the understanding the structure of the SMWDs.
In several next paragraphs, we give a brief survey of recent
developments.

Detailed study of a strongly magnetized cold electron gas and its
application to the SMWDs has recently been done in
Refs.~\cite{KM,DM,DM1,DM2}. This theory stems from the Landau
quantization of the motion of electrons in a magnetic field
\cite{LL,SDM} and of its modification to the case of a very strong
magnetic field \cite{LS} with a strength ${\cal B}\ge{\cal
B}_c$\footnote{The magnetized WDs were studied earlier for weaker
magnetic fields ${\cal B}\le{\cal B}_c$ in, e.g.,
Refs.\,\cite{BBMZ,MZBB,SUMA,WI}.}. It turns out that in systems with
small number of Landau levels, which is restricted by a suitable
choice of the strength of the magnetic field and of the maximum of
the Fermi energy $E_{\,\mathrm{F}}$ of the electron gas, the mass of
the SMWD can be in the range (2.3 - 2.6)\,$M_\odot$. It means that
the strong magnetic field can enhance the energy of the electron gas
to such a level that its pressure will force the gravity to allow
the SMWD to have a mass larger than the Chandrasekhar-Landau (CL)
limit of 1.44\,$M_\odot$ \cite{C,L}.

The existence of the WDs with the mass exceeding the CL limit has
recently been deduced from the study of the observed light curves
for highly luminous type Ia supernovae. For instance, it was found
in the case of the SN 2009dc \cite{KEAL} that such a model WD -- a
progenitor of the supernova -- can be formed, if a mass accretion is
combined with a rapid rotation. The comparison of calculations and
the observations yielded an estimate of the WD mass within the limit
of (2.2 - 2.8)\,$M_\odot$. The problem  is that although the
rotating WD is supposed to resist the gravitational collapse up to
the mass $\approx$ 2.7\,$M_\odot$ \cite{HEAL}, no convincing
calculations of such a WD mass limit exist so far in this model. The
maximum stable mass of general relativistic uniformly rotating WDs,
computed in Ref.\,\cite{BRR} within the Chandrasekhar approximation
for the equation of state, is $M_{\,\mathrm{max}}$ = 1.51595
$M_\odot$ for the average nuclear composition $\mu$ = 2. Later
calculations \cite{BRRS}, with the WD matter described by the
relativistic Feynman-Metropolis-Teller equation of state, provided
$M_{\,\mathrm{max}}$  = 1.500, 1.474, 1.467, 1.202 $M_\odot$ for
$^4$He, $^{12}$C, $^{16}$O, and $^{56}$Fe, respectively.

On the other hand, according to Refs.~\cite{DM3,DMR}, a new mass
limit $\approx$ 2.6 $M_\odot$ can be derived \cite{KM,DM,DM1} in a
model describing the SMWDs as a system of the relativistic
degenerate electron gas in the strong magnetic field, in which the
SMWD with the mass $\approx$ 2.6 $M_\odot$ lies at the end of a
mass-radius curve for the one Landau level system, corresponding to
the central magnetic field 8.8 $\times$ 10$^{17}$ G, when the Fermi
energy $\epsilon_{\,\mathrm{F}}$ in units of the electron mass is
$\epsilon_{\,\mathrm{F}}$=$E_{\,\mathrm{F}}/m_\mathrm{e}\,c^2$= 200.
To achieve such a mass, it was supposed that the continuously
accreting WD is being compressed by the gravity, which steadily
increases the strength of the magnetic field, because the total
magnetic flux is conserved. At the end, the magnetic field in the
core can exceed the critical value \mbox{${\cal B}_c$}, the WD
becomes the SMWD, and the pressure of the relativistic degenerate
electron gas is able to balance the gravity up to the point, when
the accreting mass raises the SMWD mass up to $\approx$ 2.6
$M_\odot$, after which the SMWD collapses and the supernova of the
type Ia explodes \footnote{It is interesting to notice that the
formation of a millisecond pulsar by the process of the
rotationally-delayed accretion-induced collapse of the
super-Chandrasekhar WD has been recently postulated in
Ref.~\cite{FTMSP}.}.

Let us note that the concept of the SMWDs developed in
Refs.~\cite{KM,DM,DM1,DM3,DMR} has been recently criticized by
several authors \cite{NK,CFD,CMFP,CEAL}. The response to this
criticism can be found in Refs.\,\cite{DM2,DM5}.

It should be pointed out that neither side of this dispute can
support its point of view by  consistent detailed calculations,
which should take into account violation of spherical symmetry and
effects of general relativity: rather the arguments are based on
estimates within simplified physical pictures. Any detailed analysis
of the criticism being out of the scope of this work, we would like
to support the authors and defendants of the model
\cite{KM,DM,DM1,DM3,DMR} by stressing some of the points which are,
in our opinion, not sufficiently developed in Ref.~\cite{DM2}.

The arguments presented below, together with those of the article
\cite{DM2}, justify our choice of the simple model
\cite{KM,DM,DM1,DM3,DMR} for our first estimate of the role of the
reaction (\ref{EMEPC}).

\begin{itemize}

\item  One of the arguments presented in Refs.~\cite{NK,CFD,CEAL}
is based on the statement that the SMWDs with the considered
ultra-high magnetic field should be substantially deformed, while
the model assumes the spherical symmetry.  Thus, the numerical
values of the ratio $\epsilon/R$ (surface deformation/radius),
presented in \mbox{Table I} of Ref.\,\cite{CEAL}, are calculated
from the equation
\be
\frac{\epsilon}{R}\,=\,-\frac{15}{8}\,\frac{B^2 R^4}{G M^2}\,,
\label{EOR}
\ee
where $B$ is the poloidal uniform magnetic flux density in the
z-direction, $R(M)$ is the radius (mass) of the star and $G$ is the
Newton gravitational constant. This equation was derived in
Ref.\,\cite{CF} for the case of $|\epsilon/R|\,\ll\,1$. Since the
calculated values $|\epsilon/R|\,\gg\,1$, they are questionable and
it is premature to draw from them conclusions about the shape of the
SMWDs. Another equation, similar to Eq\,(\ref{EOR}), was derived by
Ferraro \cite{VF} for the eccentricity $e_c$ and used by Bocquet et
al. \cite{BOC} in the form
\be
e_c\,=\,\frac{15}{4}\frac{B_{pole}}{\sqrt{\pi G \mu_0}\,\rho R}\,  \label{ECC}
\ee
to check the code. Here, $e_c$ is the eccentricity, computed at the
first order around the spherical symmetry. To ensure that Ferraro's
approximation is valid, Bocquet et al. considered weak enough
magnetic field $B_{pole}=6.6\times 10^5$ T and also a small star
mass $M=2.67\times 10^{-2}\, M_\odot$ of the constant density
$\rho=1.66\times 10^9$ kg/m$^{3}$, in order to have also a weak
enough gravitational field. Then it follows from Eq.\,(\ref{ECC})
that $e_c=0.04670$, quite close to the eccentricity resulting from
the code \cite{BOC}: $e_c=0.04683$. In other words, Eq.\,(\ref{ECC})
as well as Eq.\,(\ref{EOR}) can be used solely for the case of weak
gravitational and magnetic fields which cause a small deviation from
the spherical symmetry.

Moreover, as explained in detail in Ref.\,\cite{DM5} in connection
with the magnetostatic equilibrium equation (3), for the chosen
field configuration the gradient of the pressure of the magnetic
field cancels with the Lorentz force and one is left simply with the
equation for the hydrostatic equilibrium between the pressure of the
electron degenerate matter and of the gravity\footnote{Support for
this simple model, following from the results of calculations done
in more realistic model \cite{DM4}, is mentioned in connection with
the discussion of these results below.}.

Let us also note that the model calculations of the neutron stars (NSs)
\cite{BOC,PIL} do not confirm that the deformation of stars due to
even extreme magnetic fields ($B\,\sim\,10^{17} - 10^{18}$ G) is
catastrophically large. The study \cite{BOC} shows (see Fig.\,3 and
Fig.\,4) that \mbox{$(R_e-R_p)/R_p\,=\,0.18$}.
In this case, $M=1.81\,M_\odot$, $B_p\,=\,9.1\times 10^{16}$ G,
$B_c\,=\,3.57\times 10^{17}$ G and $R_e\,\approx\,18$ km. However,
in the case of the twisted-torus configuration with the poloidal
field two times weaker than the toroidal one, the spherical
symmetry is fully restored \cite{PIL}. Besides, the models of
NSs focused on the purely poloidal and purely toroidal
magnetic fields turned out to suffer from  Tayler's instability
\cite{TI}. As mentioned in \cite{CR}, the twisted-torus geometries
were studied both in the Newtonian approach and also in the approach
of the general relativity with the common result of the poloidal
field dominated geometries, which turned out to be unstable as well.
However, the very recent results show \cite{CR,RC} that one can obtain
the twisted-torus configurations where
the toroidal to total magnetic field energy ratio can be up to 90 \%
and that these toroidal field dominated configurations are good
candidates for stability.

Let us next mention the earlier important study of the magnetized
WDs within the Newtonian model with the rotation included \cite{OH}.
It was shown by Ostriker and Hartwick that a stable WD can be
constructed with the super-critical magnetic field 9.22 $\times$
10$^{13}$ G  in the center of the star and with the super-CL mass
\mbox{$M$ = 1.81 $M_\odot$} (see Model 6 in Table 1). In agreement
with the results \cite{PIL,RC,CR}, the toroidal field dominates and
the ratio of the energy of the toroidal field to the energy of the
poloidal field is 9.82. Besides, comparison of the Model 4 (without
the magnetic field) and Model 6 shows that the central matter
density is by about 2 orders larger in the super-CL Model 6, which
was overlooked by Coelho et al. \cite{CEAL}. But in the case of the
non-super-CL Model 5, the situation is reversed. It is interesting
to note that the profile of the chosen magnetic field leaves the
surface nearly spherical.

\item It is also claimed in Refs.~\cite{CFD,CMFP,CEAL} that the heavy SMWD
are unstable in respect to an inverse beta-decay, which in
particular puts an upper limit on the magnetic field in the core.

The inverse beta-decay  is usually ignored in the WDs modeling when
the electrons can be considered as non-relativistic \cite{PAB}. It
takes place in the central region of the WDs with the masses close
to the CL limit, where the Fermi energy $E_{\,\mathrm{F}}$ of the
electron gas is large enough to trigger the reaction (\ref{IBD}).
For the nucleus $^{56}_{26}$Fe, the following two-step reaction
takes place (see Ref.\,\cite{BK1}, Ch.\,5)
\begin{equation}
^{56}_{26}\mathrm{Fe}(0^+)\,\rightarrow\,^{56}_{25}\mathrm{Mn}^*(1^+)\,
\rightarrow\,^{56}_{25}\mathrm{Mn}(3^+)
\,\rightarrow\,^{56}_{24}\mathrm{Cr}(0^+)\,,       \label{R1}
\end{equation}
with $\Delta^{\beta}_{26}$=3809 keV (additional 109 keV are needed
to excite $^{56}_{25}$Mn$^*(1^+)$) and $\Delta^{\beta}_{25}$=1610
keV. The electron capture by the nucleus $^{56}_{25}$Mn$(3^+)$
proceeds in non-equilibrium and is accompanied by the heating
\cite{BK1,BS}, with an energy of 476 keV released per electron
capture. This heating essentially affects the cooling of the iron
WDs  (see \cite{BK2}, Ch.\,12) for the temperatures \mbox{$T \le$
5.5$\times$ 10$^6$ K}. Without it the full cooling of the WD from
the temperature \mbox{$T$ = 5.5$\times$ 10$^6$ K} requires 4$\times$
10$^8$ yr, but due to the non-equilibrium heating the WD cools to
$\approx$ 10$^6$ K over a cosmological time of 20 Gyr.

The inverse beta decay instability has recently been taken into
account in the calculations within the general relativity framework
in Refs.\,\cite{RRRX} and \cite{MAT}. The critical density
$\rho^{rel}_{crit}$ for the onset of the gravitational collapse,
also obtained in these calculations, differs for the iron WDs by a
factor $\approx$ 22, which illustrates the size of model dependence
in the astrophysical calculations. Besides, the calculations of
$\rho^{rel}_{crit}$ \cite{RRRX,MAT} do not take into account the
strong magnetic fields. Therefore, the comparison of the derived
$\rho^{rel}_{crit}$ \cite{RRRX} with the densities of the SMWDs made
in the last but one paragraph at p.\,5 of Ref.\,\cite{CEAL} is not
proper.

Chamel et al. \cite{CFD,CMFP} considered the inverse beta-decay in
the magnetized WDs under assumption that the gravitational collapse
of the star proceeds in equilibrium with the weak force at the pace
allowed by the rate of the reaction. In this description the
electron capture indeed seems to limit the magnetic field and matter
densities to values smaller than considered in
Refs.\,\cite{KM,DM,DM1,DM3,DMR}. However, as noted in \cite{DM2},
even these lower values of the magnetic field are large enough to
allow for the mass-radius relation approaching the super CL-limit of
the SMWDs with the mass $\ge$ 2.44 $M_\odot$.

Besides, Chamel et al. calculated the capture rates for the
oxygen-carbon configuration of the SMWD with the mass \mbox{$M
\approx $ 2.6 $M_\odot$} and with \mbox{$\gamma = {\cal B}/{\cal
B}_c $ = 2$\times$ 10$^4$}, which corresponds to the central value
of the magnetic field \mbox{${\cal B}$ = 8.8$\times$ 10$^{17}$ G}.
These calculations showed that such an SMWD would be highly unstable
against the inverse beta-decay. However, the heating from the both
reactions of the chain (\ref{R1}) was not taken into account in
Refs.\,\cite{CFD,CMFP}, which will certainly affect the cooling of
the star \cite{BK1,BS}. In this case, even the first reaction of the
chain (\ref{IBD}) proceeds in non-equilibrium and a part of the
energy of the captured electron from the range
\mbox{$\Delta^{\beta}_{Z} < E_{\,\mathrm{e}} < E_{\,\mathrm{F}}$}
dissipates in the SMWD as the heat energy, which can keep the SMWD
in a meta-stable state. This process would be similar to the
non-equilibrium matter heating during the collapse with
$E_{\,\mathrm{F}} \gg \Delta^{\beta}_{Z}$ (see \cite{BK1}, Ch.\,5).

\item

The importance of the effects of the general relativity for a star
can be estimated by a compactness parameter $x_g$ \cite{PPNS} \be
{x}_g={r}_g/{R},\quad\quad {r}_g=2{GM}/{c}^2\approx 2.95
{M}/{M}_\odot\,\,{\rm km}\,,\label{XG} \ee where $r_g$ is the
Schwarzschild radius. Using the radii from our Table \ref{tab:res1}
and the mass $M=2 M_\odot$ we obtain for the compactness parameter
$x_g$ the values \mbox{ $x_g <$ 0.03.} For the carbon SMWDs of
Refs.\,\cite{KM,DM,DM1,DM3,DMR} is \mbox{$x_g <$ 0.11.} On the other
hand, for the standard NS with $M$=1.4\,$M_\odot$ and $R$=10 km the
value of $x_g$ = 0.413. Evidently, the effects of the general
relativity should be in the SMWDs much smaller than in the NSs,
though not completely negligible, as discussed in
Refs.\,\cite{DM6,DM4}.

\end{itemize}

We hope that we have shown clearly that even after many years of
efforts the mentioned problems  are still far from being thoroughly
understood and that the criticism of the approach to the SMWDs
\cite{KM,DM,DM1,DM3,DMR} could be considered rather as a step in
continuing debate than its fundamental refusal.

Recent study aiming to shed more light on the problem of the SMWDs
has been presented in Ref.\,\cite{DM4}, where the issue of their
stability was addressed in a general relativistic framework by
adopting the magnetized Tolman-Oppenheimer-Volkoff equation given in
Ref.\,\cite{HB} for anisotropic matter described by polytropic
equations of state. In Ref.\,\cite{DM4}, this formalism was applied
for the equation of state of the form $P=K\rho^{1+1/n}$, where $n$
is the polytropic index and $K$ is the constant and for the
spherical SMWDs with the isotropic magnetic field in a hope that the
anisotropy due to the strong magnetic field would not change the
main conclusions. The mean isotropic magnetic pressure was taken as
$P_B={\cal B}^2/(24\pi)$. The solutions, presented  for the Fermi
energy $\epsilon_{\,\mathrm{F}}$ = 20 in Table 1, were obtained for
profiles of varying magnetic fields restricted by two specific
constraints of the parallel pressure. The calculations show that the
maximum stable mass of the SMWDs could be more than 3 solar masses.
The authors stress that the key point of the calculations lies in
taking into account consistently the magnetic field pressure, which
leads to higher number of Landau levels and, consequently, to lower
values of the central magnetic field. In order to avoid the problem
with the neutronization, the considered Fermi energies
$\epsilon_{\,\mathrm{F}}$ were restricted to the value
$\epsilon_{\,\mathrm{F}}$ $\le$ 50. The calculation using a very
slowly varying magnetic field provided the result close to the one
obtained earlier \cite{KM,DM,DM1}, performed under assumptions of
the spherical symmetry and of the constant magnetic field in a large
central region and of a small number of Landau levels.

Let us note that the authors \cite{DM4} also have shown that the
criticism of the work \cite{DM3} by Dong et al. \cite{DEAL} is
erroneous, because they did not include the magnetic density into
the hydrostatic equilibrium equation (for details see Eq. (1.1)
\cite{DM4}). Without it, they obtained mass $M$=24\,$M_\odot$
instead of $M$=2.58\,$M_\odot$ with the magnetic density included.

Here, a remark is in order:\\
The concept of the anisotropic matter pressure due to the magnetic
field from which stems Ref.\,\cite{HB}, was criticized in
Ref.\,\cite{PYAM}. Dexheimer et al. \cite{DPM} made an attempt to
refute this criticism for the case of the ionized Fermi gas subject
to an external magnetic field. However, as it is clear from the last
paragraph at p.\,039801-1 \cite{PYAM}, the anisotropy of the matter
pressure does not take place either in this case. This fact has been
re-discussed  at length in very recent Ref.\,\cite{NOV}, stressing
that the magnetic field does not induce any anisotropy to the matter
pressure defined thermodynamically as a derivative of the partition
function: it transforms as a scalar and it is the Lorentz force that
deforms the magnetized stars.

As we have already mentioned above, the appearance of the works
\cite{BEBH,DM6} changed the understanding of the structure of the
SMWDs essentially.

In Ref.\,\cite{BEBH}, Bera and Bhattacharya used the axisymmetric
Newtonian formalism developed earlier for studying the structure of
the rotating magnetized stars \cite{TOER,LAJO}. For the poloidal
magnetic field in the interior of the SMWD of the intensity of the
order of 10$^{14}$ G and including the effect of the Landau
quantization, Bera and Bhattacharya  obtained the maximal mass of
the star of the order of 1.9\,M$_\odot$, thus reproducing the result
gained earlier by Ostriker and Hartwick \cite{OH} without the Landau
quantization, but with the mix of the poloidal and toroidal magnetic
fields, with the prevailing toroidal component. In contrast to the
large deformation of the order of 30\,\% obtained in
Ref.\,\cite{BEBH} for the poloidal magnetic field, the type of the
magnetic field considered in \cite{OH} leaves the star nearly
spherical. As noted explicitly in \cite{BEBH}, the ability of an
SMWD to possess  more mass than is allowed by the CL limit, can be
explained by use of the Lorentz force in the basic hydrostatic
equilibrium equation. It should be also noted that, as already
discussed above, the models of stars that focused on the purely
poloidal and purely toroidal magnetic fields turned out to suffer
from Tayler's instability \cite{TI}.

In the very recent work \cite{DM6}, Das and Mukhopadhyay performed
the calculations of the properties of the SMWDs within the framework
of the general relativistic magnetohydrodynamic approach that was
earlier developed and applied to the study of strongly magnetized
neutron stars \cite{BUDZ,PIEAL}. The calculations \cite{DM6} show
that the nature of deformation induced in SMWDs, due to purely
poloidal, purely toroidal, and mixed magnetic field configurations
are similar to those obtained in strongly magnetized neutron stars
\cite{PIEAL}. In all  models considered in \cite{DM6}, the SMWDs
possessing  the super-CL mass were observed for the maximum magnetic
field strength inside the SMWDs in the range \mbox{10$^{13} \le
{\cal B}_{max} \le 10^{15}$} G, with the central density chosen as
\mbox{10$^{10} \le \rho_c \le 10^{11}$} g/cm$^3$.

Let us finally briefly mention how the SMWDs could be formed and if
there is any observational hint of their existence. It has been
recently supposed in \cite{DM3,DMR} that the SMWDs could appear as
the result of accretion. Nowadays the process of accretion is
intensively studied. So the formation of NSs via the
accretion-induced collapse of a massive WD in a close binary has
been studied recently in Ref.\,\cite{TSYL}. Detailed study of a
detached binary, SDSS J0303308.35+005444.1, containing a magnetic WD
accreting material from a main-sequence companion star via
Roche-lobe overflow has been published in Ref.\,\cite{PEAL}.

Another way of formation of the SMWDs could be binary WD mergers,
studied in another context in Refs.\, \cite{DRBP,KIM,IS,REAL}. It
was found that rather highly magnetized WDs can arise in such
mergers. It has recently been shown  \cite{MAC,MRR,CM,BIRR} that
such magnetized WDs with rather large mass $M\,\ge\,1.2\,M_\odot$,
with the magnetic field in the range $B\,\simeq\,10^7\,-\,10^{10}$ G
and with the rotating period of the order of a few seconds can
explain some properties of sources of the soft $\gamma$-ray (SGRs)
and X-ray radiation (AXPs), widely accepted as magnetars
\cite{DT,TAD}. Let us here also mention a notice on formation of
pulsar-like WDs \cite{IKB} and very recent reviews on the topics in
Refs.\,\cite{PRL,POY}.

Let us also note recent Ref.~\cite{FTMSP}, where formation of the
super-Chandrasekhar WD was supposed to proceed from the two main
sequence stars of non-equal mass via Roche-lobe mass transfer
towards the lower mass star with the subsequent conformation of a
common envelope. Subsequent envelope ejection leads to formation of
an ONeMg WD from the naked core of the heavier star, which in its
turn accrues via Roche-lobe overflow the matter from the secondary
star. As a result of accretion the WD becomes a rapidly spinning
super-Chandrasekhar WD. However, as we have already mentioned above,
there is no convincing calculation of such kind of WDs. On the other
hand, such an accretion process could cause creation of the SMWD
\cite{DM3,DMR}.

In any case, it should be perceivable that such compact objects as
SMWDs with strong magnetic field, with mass larger than the CL limit
and with small radius, are difficult to be formed. In any case, they
were not observed among the 592 magnetized WDs known at present
\cite{KKJE,KPJ}. More light on the problem can be shed soon by GAIA,
a new satellite mission of ESA \cite{MABEAL}, aiming at absolute
astrometric measurements of about one billion of stars with
unprecedent accuracy.

Having in mind the results of Ref.\,\cite{DM4}, here we estimate the
rate of the double charge exchange reaction (\ref{EMEPC}), using the
above discussed simple model of the SMWDs~\cite{KM,DM,DM1,DM3,DMR}.
The threshold for this reaction with the initial nucleus
$^{56}_{26}$Fe and the final one $^{56}_{24}$Cr is
$\Delta^{\beta\beta}_{26}$=6.33 MeV. Then for the SMIWDs with
$E_{\,\mathrm{F}}\,\ge\,\Delta^{\beta\beta}_{26}$, this reaction can
take place. We shall consider $\epsilon_{\,\mathrm{F}}$=20,\,46 and
choose the strength of the magnetic field so that the value of the
related parameter $\gamma={\cal B}/{\cal B}_c$ will allow us to
restrict ourselves to the ground Landau level.

Having calculated the rate of the reaction (\ref{EMEPC}) and the
corresponding energy production rate in the interior of the SMWDs,
we are trying to estimate, whether its influence on the effective
temperature and/or cooling of the SMIWDs could be detectable. Since
we have no data on the effective temperature or the luminosity of
the SMIWDs, in the first step our guess was that the luminosity of
the SMIWD would be in the range ($L=10^{-5}\, L_\odot$ -
$L=10^{-8}\, L_\odot$), where $L_\odot$ is the luminosity of the
Sun. In this case, we obtained that the effect of the reaction
(\ref{EMEPC}) on the surface temperature could be up to 10 \%.

To see how under the conditions described above, the energy produced
by the reaction (\ref{EMEPC}) could influence directly the cooling
of the SMIWDs, one should include this process into appropriate
detailed microscopic model. Unfortunately, such detailed models of
cooling, including all relevant modifications of the energy
transport, in particular that of the opacity of the intra-stellar
medium have not been yet elaborated for the SMIWDs. Possible way how
to do it could be looked for in a recent review \cite{AYPA} on the
current status of the theory of surface layers of NSs possessing
strong magnetic fields and of radiative processes that occur in
these layers.

In this situation, we employed the detailed results for the
luminosity of the iron-core WDs, obtained in Ref.\,\cite{PAB} and
presented at Fig.\,17. We extrapolated the corresponding data to a
region where the luminosity is comparable to its change due to the
process (\ref{EMEPC}). In this way, we qualitatively estimated that
the double charge exchange reaction (\ref{EMEPC}) could effectively
retard  cooling of the SMIWDs at low luminosity regime only, which
is out of reach of the present observational possibilities.

In analogy with Refs.\,\cite{MAC,MRR,CM,BIRR}, we have next
considered the SMIWDs as fast rotating stars, which could be the
sources of the soft $\gamma$-ray (SGR) and anomalous X-ray radiation
(AXP). For this study, we have chosen two SGRs/AXPs, namely
\mbox{SGR 0418+579} and \mbox{Swift J1822.6-1606}, for which are the
rotational period $P$, the spin-down rate $\stackrel{.}{P}$ and the
total luminosity $L_X$ well known \cite{MCG,RSGR,RPC,RSWIFT}. Our
calculations have shown that the loss of the rotational energy of
the fast rotating SWIMDs well describe the observed total luminosity
of these compact objects, like the fast rotating WDs considered in
\cite{MAC,MRR,CM,BIRR}. However, the contribution to the luminosity
from the reaction (\ref{EMEPC}) turned out to be negligibly small in
comparison with the loss of the rotational energy for this range of
the radiation.

In Section \ref{prel}, we present methods and necessary input,
needed for the calculations of the energy production due to the
reaction (\ref{EMEPC}), including the double charge exchange width
per one elementary reaction for chosen values of
$\epsilon_{\,\mathrm{F}}$ and $\gamma$. In Section \ref{EPIWD}, we
give our estimate of the energy production and its contribution to
the luminosity.  Further, in Section \ref{cwd}, we compare our
results with the luminosity of the iron-core WDs, studied by Panei
et al. \cite{PAB}. Then in Section \ref{sgr_axp}, considering the
SMIWDs as fast rotating stars, we calculate the total luminosity as
the loss of the rotational energy. For this model, we take the input
data as observed for the magnetars \mbox{SGR 0418+579} and
\mbox{Swift J1822.6-1606}.

In Section \ref{concl}, we discuss our results and present our
conclusions. Finally, invariant functions, entering the
positron-electron annihilation probability, are defined in Appendix
\ref{appA} .

Our main conclusion is that the energy, released in reaction
(\ref{EMEPC}), (i) could influence under certain assumption the
effective temperature up to 10  \%, but the SWIMDs would be too dim
to be observed at present; (ii) could influence cooling of the
SMIWDs at sufficiently low luminosity, which seem to be, however, at
an unobservable level at present as well; (iii) cannot influence
sizeably the luminosity of these compact objects considered as the
sources of SGR/AXP radiation. It means that the study of the double
charge exchange reaction (\ref{EMEPC}) in the SMIWDs using simple
model \cite{KM,DM,DM1,DM3,DMR} with the ground Landau level and at
the present level of accuracy of measurement of the luminosity and
energy of the cosmic gamma-rays could not provide conclusive
information on the Majorana nature of the neutrino, if its effective
mass would be $|\langle m_\nu \rangle|$ $\le$ 0.8 $eV$.

From now on, we use the natural system of units with $\hbar=c=1$.

\section{Methods and inputs}
\label{prel}

In this section, we first discuss the necessary ingredients of the
theory of the SMWDs and then we present the formalism for the
calculations of the capture rate for the double charge exchange
reaction (\ref{EMEPC}).

\subsection{Theory of strongly magnetized white dwarfs}
\label{thmwd}

Theory of SMWDs is based on the Landau quantization of the motion of
free electrons in a homogenous magnetic field ${\cal B}$, usually
taken to point along the z-axis \cite{LL,SDM}. It is discussed in
detail in Refs.\,\cite{LS,KM,DM,DM1,DM2}. In the relativistic case,
one solves the Dirac equation, obtaining for the electron energy
\be E_\nu=m_\mathrm{e}\, \left[1 + \left(\frac{p_z}{m_\mathrm{e}}\right)^2 + 2\nu\,
\frac{e\, {\cal B}}{m^2_\mathrm{e}  } \, \right]^{1/2}\  . \label{ENU} \ee
Here, $e$ is the electron charge, $p_z$ is the electron momentum
along the z-axis and $\nu=l+1/2+\sigma$ labels the Landau levels
with the principal number $l$, $\sigma=\pm 1/2$. The ground level
($\nu$=0) is obtained for $l$=0 and $\sigma$=-1/2, and it has the
degeneracy 1. Other Landau levels possess the degeneracy 2.

The last term at the r.h.s. of Eq.\,(\ref{ENU}) can be written as
\be 2\nu\, \frac{e\, {\cal B}}{m^2_\mathrm{e}  }= 2\nu\, \frac{
\omega_\mathrm{H}}{m_\mathrm{e} }= 2\nu\,  \frac{\cal B}{{\cal
B}_c}= 2\nu\,\gamma . \label{REL} \ee
Here, the cyclotron frequency $\omega_\mathrm{H}= e\, {\cal
B}/m_\mathrm{e}$ and a critical magnetic field strength
$$
{\cal B}_c=\frac{m^2_\mathrm{e} }{ e}=4.414 \times 10^{13}\
{\rm G} \ .
$$
One can see from Eqs.\,(\ref{ENU}) and (\ref{REL}) that the electron
becomes relativistic if $\omega_\mathrm{H} \ge m_\mathrm{e}$, or if
${\cal B}\ge{\cal B}_c$.

In contrast to the  density of electron states in the absence of the
strong magnetic field, given as $2\, d^3p/(2\pi)^3$, the
presence of such magnetic field modifies the number of electron
states for a given level $\nu$ to $ 2 g_\nu\, e\, {\cal B}\, dp_z/(2\pi)^2$.
Then the sum over the electron states
in the presence of the strong magnetic field is given by
\be \sum_E \ \rightarrow \ \sum_\nu \frac{2 e{\cal B}}{(2\pi)^2 }\,
g_\nu\, \int\, dp_z =\frac{2\gamma}{(2\pi)^2\lambda^3_\mathrm{e}}\,
\sum_\nu g_\nu\, \int d\left(\frac{p_z}{m_\mathrm{e} }\right)\ .
\label{SOES} \ee
Here, $\lambda_\mathrm{e}= 1/m_\mathrm{e} $ is the electron Compton
wavelength and $g_\nu=(2-\delta_{0,\nu})$ reflects the degeneracy of
the Landau levels.

The relation between the Fermi energy $\epsilon_{\,\mathrm{F}}$ and
the Fermi momentum $x_{\,\mathrm{F}}(\nu)=p_{\,\mathrm{F}}(\nu)/m_\mathrm{e}$
for the Landau level, specified by $\nu$, is obtained directly from
Eq.\,(\ref{ENU}):
\be
\epsilon^2_{\,\mathrm{F}}=x^2_{\,\mathrm{F}}(\nu)+(1+2\nu\gamma)\ .
\label{EFPF}
\ee
Then the following equation for the electron number density follows
from Eq.\,(\ref{SOES})
\be n_{\mathrm{e}^-}=\frac{2\gamma}{(2\pi)^2\lambda^3_\mathrm{e}}\,
\sum^{\nu_{max}}_0\, g_\nu\, x_{\,\mathrm{F}}(\nu)\ .  \label{NE}
\ee
The values of $\nu$ in the sum are restricted by the condition
$x_{\,\mathrm{F}}(\nu)\ge 0$, implying
$\epsilon^2_{\,\mathrm{F}}-(1+2\nu\gamma)\ge 0$, from which the
inequality follows
\be
\nu_{max}\le integer\left(\frac{\epsilon^2_{{\,\mathrm{F}}
max}-1}{2\gamma}\right)\,.   \label{NUM}
\ee
Assuming that the WDs are electrically neutral, one deduces  the
matter density for the system of the electron gas and one sort of
nuclei
\be \rho_\mathrm{m}= \mu_{\mathrm{e}^-} m_\mathrm{U}
n_{\mathrm{e}^-}=\frac{n_{\mathrm{e}^-}}{Z} m_A\,, \label{RHOM} \ee
where $\mu_{\mathrm{e}^-}=A/Z$ is the molecular weight per electron
[$A(Z)$ is the mass (atomic) number of the nucleus], $m_\mathrm{U}$
is the atomic mass unit and $m_A$ is the mass of the nucleus with
the mass number $A$. For the lightest nuclei,
$\mu_{\mathrm{e}^-}=2$, but for $^{56}_{26}$\,Fe, e.g., one obtains
$\mu_{\mathrm{e}^-}=2.15$.

The electron energy density at zero temperature reads
\bea
\varepsilon_{\mathrm{e}^-}&=&\frac{2\gamma}{(2\pi)^2\lambda^3_\mathrm{e}}\,
\sum\limits_0^{\nu_{max}}
\, g_\nu\,\int\limits_0^{x_{\,\mathrm{F}}}\,E_\nu dx(\nu) \nonumber\\
 &=&\frac{2\gamma m_\mathrm{e}}{(2\pi)^2\lambda^3_\mathrm{e}}\,\sum_0^{\nu_{max}}
 \,g_\nu\,\int\limits_0^{x_{\,\mathrm{F}}}\,
\left[1+x^2(\nu)+2\nu\gamma\right]^{1/2}\,dx(\nu) \nonumber\\
&=&\frac{\gamma
m_\mathrm{e}}{(2\pi)^2\lambda^3_\mathrm{e}}\,\sum_0^{\nu_{max}}
\,g_\nu\,\left[x_F(\nu)\, \epsilon_{\,\mathrm{F}}+(1+2\nu\gamma)\,
\ln\frac{x_{\,\mathrm{F}}(\nu)+\epsilon_F}
{\left(1+2\nu\gamma\right)^{1/2}}\right]\  , \label{ENUT} \eea
the energy per electron is then
\be
\bar{\varepsilon}_{\mathrm{e}^-}=\varepsilon_{\mathrm{e}^-}/n_{\mathrm{e}^-}\,.
\label{MEE} \ee
One can see  from Eqs.\,(\ref{NE}), (\ref{ENUT}) and (\ref{MEE})
that for the ground Landau level
$\bar{\varepsilon}_{\mathrm{e}^-}=E_{\,\mathrm{F}}/2$.

The pressure of the degenerate electron gas, related to the energy
density (\ref{ENUT}) is
\bea
P_{\mathrm{e}^-}&=&\frac{2\gamma}{(2\pi)^2\lambda^3_\mathrm{e}}\,\sum_0^{\nu_{max}}
\,g_\nu\,\int\limits_0^{x_{\,\mathrm{F}}}\,
\frac{x^2(\nu)}{\left[1+x^2(\nu)+2\nu\gamma\right]^{1/2}}
\, dx(\nu) \nonumber\\
 &=&\frac{\gamma m_\mathrm{e}}{(2\pi)^2\lambda^3_\mathrm{e}}\sum_0^{\nu_{max}}
\,g_\nu\,\left[x_F(\nu)\, \epsilon_{\,\mathrm{F}}-(1+2\nu\gamma)\,
\ln\frac{x_{\,\mathrm{F}}(\nu)+\epsilon_F}
{\left(1+2\nu\gamma\right)^{1/2}}\right]\,. \label{PNUT}
\eea

Here, we restrict ourselves to the SMIWDs with the electrons,
occupying the ground Landau level ($\nu_{max}=0$) and consider
$\epsilon_{\,\mathrm{F}}$= 20 and 46. We choose the parameter
$\gamma$ to be $\gamma$=$\epsilon^2_{\,\mathrm{F}}/2$, which is the
minimal value of $\gamma$ satisfying Eq.\,(\ref{NUM}). In
Table~\ref{tab:inpt}, we present quantities needed for the
calculations.

For the ground Landau level, the electron pressure can be written in
the polytropic form \be
P_{\mathrm{e}^-}\,=\,K\,\rho_\mathrm{m}^{\Gamma}\,, \label{PPF} \ee
where \be K\,=\,\frac{\pi^2 }{m_\mathrm{e}^2 \,
m^2_\mathrm{U}}\,\frac{1}{\gamma \mu^2_\mathrm{e}} \,=\,\frac{1.686
\times 10^{11}}{\gamma \mu^2_\mathrm{e}}\quad \mathrm{cgs}.\ ,
\label{CK} \ee and \be \Gamma\,=\,1+\frac{1}{n}\,=\,2\ , \label{GAM}
\ee from which one obtains the value of the polytropic index $n=1$.

\begin{table}[tb]
\caption{The values of the Fermi energy
$\epsilon_{\,\mathrm{F}}$=$E_{\,\mathrm{F}}/m_\mathrm{e}$, used in
the present study. Further, $n_{\mathrm{e}^-}$ is the electron
number density, $\rho_{\mathrm{e}^-}$ is the corresponding electron
density, $\rho_\mathrm{m}$ is the matter density, calculated
according to Eq.\,(\ref{RHOM}) for the nuclei $^{56}_{26}$\,Fe, and
the values of $\gamma$ are the smallest values, satisfying
Eq.\,(\ref{NUM}).}
\begin{center}
\begin{tabular}{|l |  c    | c    |  c    |  c    |}\hline
$\epsilon_{\,\mathrm{F}}$ & $n_{\mathrm{e}^-}$/10$^{33}$ [1/cm$^3$]
&$\rho_{\mathrm{e}^-}$/10$^6$ [g/cm$^3]$  &
$\rho_\mathrm{m}$/10$^{10}$ [g/cm$^3]$ & 2$\gamma$ \\\hline
 20        & 3.52 & 3.20 & 1.26 & 400   \\
 46        & 42.8 & 13.4 & 15.2 & 2116  \\\hline
\end{tabular}
 \end{center}
\label{tab:inpt}
\end{table}

\subsection{Reaction rate}
\label{reacrate}

The process of $(\mathrm{e}^-,\mathrm{e}^+)$ conversion
(\ref{EMEPC}) is very similar to the neutrinoless double beta decay
($0\nu\beta\beta$-decay)
\begin{equation}
\mathrm{X}(A,Z) \rightarrow \mathrm{X}(A,Z+2) + \mathrm{e}^- +
\mathrm{e}^-\,.
\end{equation}
Both processes violate total lepton number by two units and
therefore take place if and only if neutrinos are Majorana particles
with the non-zero mass. Moreover, as we will show below, the
$(\mathrm{e}^-,\mathrm{e}^+)$ conversion rate is, like the
$0\nu\beta\beta$-decay rate, proportional to the squared absolute
value of the effective mass of Majorana neutrinos $|\langle
m_\nu\rangle|^2$. This quantity is defined as
\begin{equation}
\langle  m_\nu \rangle =  \sum_{i=1}^{3} U^2_{ei}\, m_i\, .   \label{MBB}
\end{equation}
Here, $U$ is the $3\times3$ Pontecorvo-Maki-Nakagawa-Sakata unitary
mixing matrix and $m_i$ ($i=1,2,3$) is the mass of the i-th light
neutrino. Let us note that $\langle m_\nu \rangle$ depends on
neutrino oscillation parameters $\theta_{12}$, $\theta_{13}$,
$\Delta m^2_{\mbox{\tiny{SUN}}}$, $\Delta m^2_{\mbox{\tiny{ATM}}}$,
the lightest neutrino mass and the type of the neutrino mass
spectrum (normal or inverted).

From the most precise  experiments on the search for the
$0\nu\beta\beta$-decay \cite{bau99,te130,kamlandzen} the following
stringent bounds were inferred \cite{VES}:
\begin{eqnarray}\label{bounds}
|\langle m_\nu\rangle| & < & (0.20-0.32) ~\mathrm{eV}~~ (^{76}\mathrm{Ge}), \nonumber\\
               & < & (0.33-0.46) ~\mathrm{eV}~~ (^{130}\mathrm{Te}), \nonumber\\
               & < & (0.17-0.30) ~\mathrm{eV}~~ (^{136}\mathrm{Xe}) \ ,
\end{eqnarray}
by use of nuclear matrix elements (NME) of Ref. \cite{src09}.
However, there exists a claim of the observation of the
$0\nu\beta\beta$-decay of $^{76}\rm{Ge}$, made by some participants
of the Heidelberg-Moscow collaboration \cite{evidence2}. Their
estimated value of the effective Majorana mass (assuming a specific
value for the NME) is $|\langle m_\nu\rangle|\simeq 0.4$ eV. In
future experiments, CUORE (${^{130}\rm{Te}}$), EXO, KamLAND-Zen
(${^{136}\rm{Xe}}$), MAJORANA/GERDA (${^{76}\rm{Ge}}$), SuperNEMO
(${^{82}\rm{Se}}$), SNO+ (${^{150}\rm{Nd}}$), and others \cite{AEE08}, a
sensitivity
\begin{equation}\label{sensitiv}
|\langle m_\nu\rangle|\simeq \mathrm{a~few}~10^{-2}~\mathrm{eV}
\end{equation}
is planned, which is the region of the inverted hierarchy of
neutrino masses.  In the case of the normal mass hierarchy,
$|\langle m_\nu\rangle|$ is too small to be probed in the
$0\nu\beta\beta$-decay experiments of the next generation.

For the sake of simplicity, the $(\mathrm{e}^-,\mathrm{e}^+)$
conversion on nuclei is considered only for the ground state to
ground state transition, which is assumed to give the dominant
contribution. The spin and parity of initial (${^{56}\rm{Fe}}$) and
final (${^{56}\rm{Cr}}$) nuclei in the ground state are equal,
namely $0^+$. The incoming electron and outgoing positron are
considered to be presumably in the $s_{1/2}$ wave-states \cite{doi}
\begin{eqnarray}\label{WFS1e}
\psi^{(s_{1/2})}_{\rm{e}^-}(P_{\rm{e}^-}) &\approx&
\sqrt{F_0(Z,E_{\rm{e}^-})}~ u(P_{\rm{e}^-}),
\\
\label{WFnu} \psi^{(s_{1/2})}_{\rm{e}^+}(P_{\rm{e}^+}) &\approx&
\sqrt{F_0(Z-2,E_{\rm{e}^+})}~ u(P_{\rm{e}^+}).
\end{eqnarray}
The Coulomb  interaction of electron and positron with the nucleus
is taken into account by the relativistic Fermi functions
$F(Z,E_{\rm{e}^-})$ and $F(Z-2,E_{\rm{e}^+})$ \cite{doi},
respectively. We use the non-relativistic normalization of spinors:
$u^\dagger(P)u(P) = 1$ and $v^\dagger(P)v(P) = 1$. The 4-momenta of
electron and positron are $P_{\rm{e}^-}\equiv
(E_{\rm{e}^-},{\mathbf{p}}_{\rm{e}^-})$ and $P_{\rm{e}^+}\equiv
(E_{\rm{e}^+},{\mathbf{p}}_{\rm{e}^+})$, respectively. The above
approximation is expected to work reasonably well for the energy of
incoming electron below 50 MeV.

The leading order $(\rm{e}^-,\rm{e}^+)$ conversion matrix element
reads
\begin{eqnarray}
  \label{S-matrix}
    \langle f \vert S^{(2)} \vert i \rangle =     2 \pi \delta(E_{\rm{e}^+}-E_{\rm{e}^-}
    + E_f - E_i)
     \langle f \vert T^{(2)} \vert i \rangle\,,
\end{eqnarray}
with
\begin{eqnarray}
 \label{T-matrix}
\langle f \vert T^{(2)} \vert i \rangle &=&  \mathrm{i} ~\langle
m_\nu\rangle^* ~ \frac{1}{4 \pi} G^2_\beta
\sqrt{F_0(Z,E_{\rm{e}^-})} \sqrt{F_0(Z-2,E_{\rm{e}^+})}~
\overline{v}(P_{\rm{e}^+}) (1 + \gamma_5) u(P_{\rm{e}^-})\times \nonumber\\
&&  \frac{ g^2_{\mathrm{A}}}{R} {M}^{(\rm{e}\beta^+)}.
\end{eqnarray}
Here, $G_\beta= G_{\,\mathrm{F}}\cos\theta_c$ and $E_i$ ($E_f$) is
the energy of the initial (final) nuclear ground state. Later on, we
neglect the kinetic energy of the final nucleus. The conventional
normalization factor of the NME  ${M}^{(\rm{e}\beta^+)}$, presented
in Eqs.\,(\ref{F-light}) and (\ref{GT-light}), involves  the nuclear
radius $R =1.2~A^{1/3}~{\mathrm{fm}}$. For the weak axial coupling
constant $g_\mathrm{A}$, we adopt the value $g_\mathrm{A}=1.269$.

The NME in Eq.~(\ref{T-matrix}) is a sum of the Fermi
$M^{(\rm{e}\beta{+})}_{\,\mathrm{F}}$  and the Gamow-Teller
$M^{(\rm{e}\beta^{+})}_{\mathrm{GT}}$  contributions:
\begin{equation}
  {M}^{(\rm{e}\beta^+)} =
  -\frac{M^{(\rm{e}\beta^+)}_{\,\mathrm{F}}}{g_{\mathrm{A}}^2}
  + M^{(e\beta^+)}_{\mathrm{GT}} \, .  \label{MEBP}
\end{equation}
They take the following form
\begin{eqnarray}\label{F-light}
    M^{(\rm{e}\beta^{+})}_{\,\mathrm{F}} &=& \frac{4 \pi  R }{(2 \pi)^3}
    \int \frac{d \bm{q}}{2 q} f^2_{\mathrm{V}}({q}^{2})\times
    \nonumber\\
    &&~ \sum_n \left(
    \frac{%
    \langle 0^+_f \vert \sum_l \tau^+_l
    \mathrm{e}^{-\mathrm{i} \bm{q} \cdot \bm{r}_l}
    \vert n \rangle
    \langle n \vert \sum_m \tau^+_m
    \mathrm{e}^{\mathrm{i} \bm{q}\cdot \bm{r}_m}
    \vert 0^+_i \rangle}
    {q - E_{\rm{e}^-} + E_n - E_i + \mathrm{i} \varepsilon_n} \right.\nonumber\\
    && ~~~~+ \left.
    \frac{%
    \langle 0^+_f \vert \sum_m \tau^+_m
    \mathrm{e}^{\mathrm{i} \bm{q} \cdot \bm{r}_m}
    \vert n \rangle \langle n \vert \sum_l \tau^+_l
    \mathrm{e}^{-\mathrm{i} \bm{q} \cdot \bm{r}_l}
    \vert 0^+_i \rangle}
    {q + E_{\rm{e}^+} + E_n - E_i + \mathrm{i} \varepsilon_n}
    \right),
\end{eqnarray}
\begin{eqnarray}\label{GT-light}
    M^{(\rm{e}\beta^+)}_{\mathrm{GT}} &=& \frac{4 \pi  R }{(2 \pi)^3}
    \int \frac{d\bm{q}}{2 q} f^2_{\mathrm{A}}({q}^{2})\times
    \nonumber\\
    &&~ \sum_n \left(
    \frac{%
    \langle 0^+_f \vert \sum_l \tau^+_l \bm{\sigma}_l
    \mathrm{e}^{-\mathrm{i} \bm{q} \cdot \bm{r}_l}
    \vert n \rangle \cdot
    \langle n \vert \sum_m \tau^+_m \bm{\sigma}_m
    \mathrm{e}^{\mathrm{i} \bm{q} \cdot \bm{r}_m}
    \vert 0^+_i \rangle}
    {q - E_{\rm{e}^-} + E_n - E_i + \mathrm{i} \varepsilon_n} \right.\nonumber \\
    &&~~+ \left.
    \frac{%
    \langle 0^+_f \vert \sum_m \tau^+_m \bm{\sigma}_m
    \mathrm{e}^{\mathrm{i} \bm{q} \cdot \bm{r}_m}
    \vert n \rangle \cdot
    \langle n \vert \sum_l \tau^+_l \bm{\sigma}_l
    \mathrm{e}^{-\mathrm{i} \bm{q} \cdot \bm{r}_l}
    \vert 0^+_i \rangle}
    {q + E_{\rm{e}^+} + E_n - E_i + \mathrm{i} \varepsilon_n}
    \right).
\end{eqnarray}
We use the conventional dipole parametrization for the nucleon form
factors, normalized to unity
\begin{equation}\label{dipole}
    f_{\mathrm{V}}({q}^{\;2}) =
    \left(1+\frac{{q}^{\;2}}{M_{\mathrm{V}}^2} \right)^{-2},~~~~~~
    f_{\mathrm{A}}({q}^{\;2}) =
    \left(1+\frac{{q}^{\;2}}{M_{\mathrm{A}}^2} \right)^{-2},
\end{equation}
with $M_{\mathrm{V}} = 0.71~\mathrm{GeV}$ and $M_{\mathrm{A}} =
1.091~\mathrm{GeV}$. In the denominators of Eqs. (\ref{F-light}) and
(\ref{GT-light}), $E_n$ and $\varepsilon_n$ are the energy and width
of the n-th intermediate nuclear state, respectively.

For the considered $(\mathrm{e}^-,\mathrm{e}^+)$ conversion on
${^{56}\mathrm{Fe}}$, the typical momentum of intermediate neutrinos
is about 200 MeV (as in the case of the $0\nu\beta\beta$-decay
\cite{anatomy}), i.e. significantly larger than  typical excitation
energies of the intermediate nuclear states.  Thus, in Eqs.
(\ref{F-light}) and (\ref{GT-light}), we complete the sum over the
virtual intermediate nuclear states by closure,  replacing $E_n -
E_i$ and $\varepsilon_n$ with some average values $\langle E_n -E_i
\rangle$ and $\varepsilon$, respectively
\begin{eqnarray} \sum_n
\frac{\vert n \rangle \langle n \vert}
      {q - E_{\rm{e}^-} + E_n - E_i + \mathrm{i} \varepsilon_n}
      &\approx&
      \frac{1}{q - E_{\rm{e}^-} + \langle E_n - E_i \rangle
      + \mathrm{i} \varepsilon}, \nonumber \\
      \sum_n \frac{\vert n \rangle \langle n \vert}
      {q + E_{\rm{e}^+} + E_n - E_i + \mathrm{i} \varepsilon_n}
      &\approx &
      \frac{1}{q + E_{\rm{e}^+} + \langle E_n - E_i \rangle
      + \mathrm{i} \varepsilon}.
\end{eqnarray}
As a result, the nuclear matrix element ${M}_{\nu}^{(\mathrm{e}^-
\mathrm{e}^{+})}$, decomposed into the contributions coming from
direct and cross Feynman diagrams,  takes the form
\begin{eqnarray}\label{eq:MEL}
{M}^{(\rm{e}\beta^{+})} = M_{\mathrm{dir.}}^{(\rm{e}\beta^{+})} +
M_{\mathrm{cro.}}^{(\rm{e}\beta^{+})},
\end{eqnarray}
with
\begin{eqnarray}
  \label{eq:Mdir}
    M_{\mathrm{dir.}}^{(\rm{e}\beta^{+})} &=&
    \langle 0^+_i \vert \sum_{lm} \tau^+_l \tau^+_m\,
    \frac{R}{\pi}\, \int\limits_0^\infty
    \frac{\mathrm{j}_0(q r_{lm}) f^2(q^2)\, q dq}
    {q - E_{\rm{e}^-} + \langle E_n - E_i \rangle + \mathrm{i} \varepsilon}
    \left(\bm{\sigma_l} \cdot \bm{\sigma_m} -
    \frac{1}{g^2_{\mathrm{A}}}\right) \vert 0^+_f \rangle \ , \\
  \label{eq:Mcro}
    M_{\mathrm{cro.}}^{(\rm{e}\beta^{+})} &=&
    \langle 0^+_i \vert \sum_{lm} \tau^+_l \tau^+_m\,
    \frac{R}{\pi}\, \int\limits_0^\infty
    \frac{\mathrm{j}_0(q r_{lm}) f^2(q^2)\, q dq}
    {q + E_{\rm{e}^+} + \langle E_n - E_i \rangle + \mathrm{i} \varepsilon}
    \left(\bm{\sigma_l} \cdot \bm{\sigma_m} -
    \frac{1}{g^2_{\mathrm{A}}}\right)
    \vert 0^+_f \rangle \ .
\end{eqnarray}
It is important to note that the value of  $E_{r} \equiv
-E_{\mathrm{e}^{-}} + <E_{n} - E_{i}>$ is negative for considered
values of $E_{\mathrm{e}^-}$ and the studied nuclear system $A$ =
56. Therefore, the contribution of direct Feynman diagram with the
light intermediate neutrino has the pole at $q=-E_r - i\varepsilon$,
as it follows from Eq.\,(\ref{eq:Mdir}). Therefore, there is a
non-zero imaginary part of the $(e{-}, e^{+})$ conversion amplitude,
which has to be properly included.

The following comment is in order. In Eq.\,(\ref{MEBP}) for the
nuclear matrix elements $M^{(e\beta^{+})}$, we neglect the
contributions of the higher order terms of the nucleon current
(weak-magnetism, induced pseudoscalar coupling). As suggested by the
analogy with $0\nu\beta\beta$-decay, these terms should be less
important for the light neutrino exchange mechanism.

Now we are ready to write down the expression for
g.s.\,$\rightarrow$\, g.s. (e$^{-}$,e$^{+}$) conversion rate. The
differential capture rate can be written as
\begin{eqnarray}
d\Gamma^{(e\beta^+)} = \sum |<f|T^{(2)}|i>|^2\, ~2\pi \delta(E_{e^-} +
E_i - E_f - E_{e^+})\, ~\frac{d{\mathbf{p}}_{e^+}~V}{(2\pi)^3} \ ,
\label{DG1}
\end{eqnarray}
where $V$ is a volume of the phase space. Calculating the modulus of
the squared T-matrix element, averaging it over the spin projections
of the initial particles and summing over the spin projections of
the final particles, we get
\begin{eqnarray}
\Gamma^{(e\beta^+)} =  |\langle m_\nu\rangle|^2  \frac{1}{V} \frac{1}{16
\pi^3}
{\left(\frac{G_{\beta}}{~\sqrt{2}}\right)}^{{4}} F_0(Z,E_{e^-})
F_0(Z-2,E_{e^+})\, \frac{g^4_A}{R^2}\left|{M}^{(e\beta^+)}\right|^2\,
p_{e^+}~E_{e^+}\  .    \label{RR}
\end{eqnarray}
Here,, $E_{e^+} = E_{e^-} - \Delta^{\beta\beta}_{56}$.

The  result obtained above is for a single electron in the volume
$V$. Assuming the density of the electrons $n_{e^-}$ (we replace
$1/V$ with $n_{e^-}$), the reaction rate per nucleus is
\begin{eqnarray}
\Gamma^{(e\beta^+)} & =& m_\mathrm{e}~ \frac{|\langle m_\nu\rangle|^2}{m^2_\mathrm{e}} ~
\frac{1}{16 \pi^3}
{\left(\frac{G_{\beta} m^2_\mathrm{e}}{~\sqrt{2}}\right)}^{{4}} ~F_0(Z,E_{e^-})
~F_0(Z-2,E_{e^+})  \nonumber \\
&& ~\times~ \frac{g^4_A}{(R^2 m_\mathrm{e}^2)}~\left|{M}^{(e\beta^+)}\right|^2
 ~\frac{p_{e^+}~E_{e^+}}{m^2_\mathrm{e}}~
\frac{n_{e^-}}{m^3_\mathrm{e}} \  .   \label{RRPN}
\end{eqnarray}

For the reaction (\ref{EMEPC}), occurring in the SMWDs, one should
sum up the rate (\ref{RRPN}) over the electron energies according to
Eq.\,(\ref{SOES}). This summation implies a replacement:
\be F_0(Z,E_{e^-})~F_0(Z-2,E_{e^+})~\frac{p_{e^+}~E_{e^+}}{m^2_\mathrm{e}}~
\frac{n_{e^-}}{m^3_\mathrm{e}}\ \rightarrow \
\phi(\epsilon_{\,\mathrm{F}},\gamma)\ , \label{CH}
\ee
where the function $\phi(\epsilon_{\,\mathrm{F}},\gamma)$ is defined
as
\bea
\phi(\epsilon_{\,\mathrm{F}},\gamma)&=&\frac{2\gamma}{(2\pi)^2\lambda^3_e
m^3_\mathrm{e}}\, \int\limits_{Q+1}^{\epsilon_{\,\mathrm{F}}}\,
\left[\frac{(\epsilon_{e^-}-Q)^2-1}{\epsilon_{e^-}-1}\right]^{1/2}
\,(\epsilon_{e^-}-Q)\,\epsilon_{e^-} \nonumber \\
&&\qquad \qquad \qquad \times F_0(Z,\epsilon_{e^-})~F_0(Z-2,\epsilon_{e^+})\,
d\epsilon_{e^-}\ . \label{ffi}
\eea
Here, $\epsilon_{e^\pm}=E_{e^\pm}/m_\mathrm{e} c^2$ and Q=$\Delta^{\beta\beta}_{56}/m_\mathrm{e} c^2$.
The calculated values of $\phi$ for chosen $\epsilon_{\,\mathrm{F}}$
and $\gamma$ are
\bea
\phi(20,200)&=&1.80\,\times\,10^3\,,   \nonumber \\
\phi(46,1058)&=&7.94\,\times\,10^5\,.   \label{vfi} \eea
Using this notation, the reaction rate in the SMIWDs reads:
\bea \Gamma^{(\mathrm{e}\beta^+)} & =& m_\mathrm{e}~ \frac{|\langle
m_\nu\rangle|^2}{m^2_\mathrm{e}} ~ \frac{1}{16 \pi^3}
{\left(\frac{G_{\beta}\, m^2_\mathrm{e}}{~\sqrt{2}}\right)}^{{4}}
{\left(\frac{g^2_\mathrm{A}}{R\,
m_\mathrm{e}}\right)}^2~\left|{M}^{(\mathrm{e}\beta^+)}\right|^2
\phi(\epsilon_{\,\mathrm{F}},\gamma)\,.  \label{RRPNF} \eea

\subsection{Nuclear matrix element}
\label{nme}

To calculate nuclear matrix element for the transition
$(\mathrm{e}^-,\,\mathrm{e}^+)$ on $^{56}$Fe, we use the
Quasiparticle Random Phase Approximation (QRPA)
\cite{anatomy,src09}. For the $A$=56 system, the single-particle
model space consisted of $0-4\hbar\omega$ oscillator shells, both
for the protons and the neutrons. The single particle energies are
obtained by using a Coulomb--corrected Woods--Saxon potential. We
derive the two-body $G$-matrix elements from the Charge Dependent
Bonn one-boson exchange potential \cite{CDB} within the Brueckner
theory. The pairing interactions are adjusted to fit the empirical
pairing gaps.

The particle-particle and particle-hole channels of the $G$-matrix
interaction of the nuclear Hamiltonian are renormalized by
introducing the parameters $g_\mathrm{pp}$ and $g_\mathrm{ph}$,
respectively. The calculations are carried out for $g_\mathrm{ph} =
1.0$. The particle-particle strength parameter $g_\mathrm{pp}$ of
the QRPA is fixed by the assumption that the matrix element
$M^{\nu\overline\nu}_\mathrm{GT}$ of  the lepton number conserving
process of electron to positron conversion on nuclei with emission
of neutrino and antineutrino
\begin{equation}
\mathrm{e}^- + {\mathrm X}(A,Z) \rightarrow {\mathrm X}(A,Z-2) +
\mathrm{e}^+ + \nu_\mathrm{e} + {\overline\nu_\mathrm{e}} \
\end{equation}
is within the range $(0,\,0.30)~$MeV$^{-1}$. Recall that a
comparable quantity $M^{2\nu}_\mathrm{GT}$, associated with the
two-neutrino double beta decay of $^{48}$Se, $^{76}$Ge, $^{82}$Se,
$^{128,130}$Te and $^{136}$Xe, does not exceed the above range by
assuming the weak-axial coupling constant $g_\mathrm{A}$ to be
unquenched ($g_\mathrm{A} = 1.269$) or quenched ($g_\mathrm{A} =
1.0$).

As we already commented above, the matrix element
$M_{\mathrm{dir.}}^{(\mathrm{e}\beta^{+})}$ (see Eq.(\ref{eq:Mdir}))
of the direct contribution contains an imaginary part  from the pole
of the integrand at $q = -E_r - i\varepsilon$. The averaged energy
of the intermediate nuclear states $\langle E_n - E_i \rangle$ is
assumed to be 5 MeV. Taking into account that the widths of
low-lying nuclear states are negligible in comparison to their
energies, one can separate the imaginary and the real parts of this
matrix element using the well-known formula
\begin{equation}
\frac{1}{\alpha + i \varepsilon} = {\cal P}\frac{1}{\alpha}
 - i \pi \delta(\alpha)\ ,
\end{equation}
valid in the limit $\varepsilon \rightarrow 0$.

\begin{figure}[!t]
\includegraphics[scale=.52]{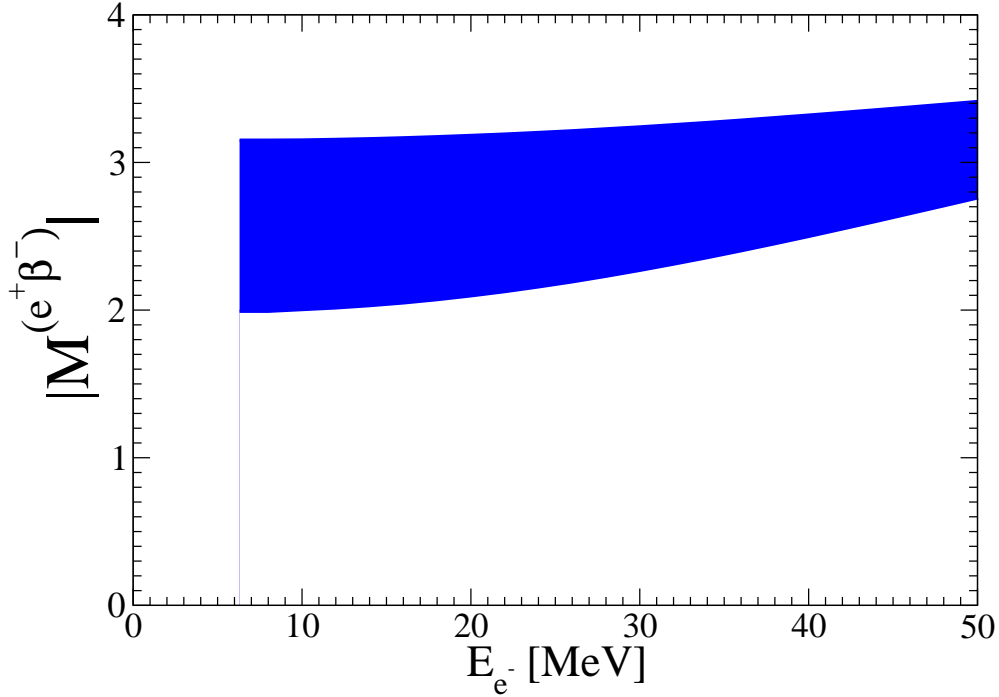}
 \vspace*{+1.0cm}
\caption{(Color online) The absolute value of nuclear matrix element
${M}^{(\mathrm{e}\beta^{+})}$ for $^{56}$Fe as function of energy of
incoming electron $E_{\mathrm{e}^-}$. The allowed region of
calculated nuclear matrix element is determined by fixing the
particle--particle parameter to $M^{\nu\overline\nu}_\mathrm{GT}$
from the range (0,\,0.30)~MeV$^{-1}$. \label{fig.1}}
\end{figure}

In Fig.~\ref{fig.1}, the absolute value of the nuclear matrix
element ${M}^{(\mathrm{e}\beta^{+})}$ for $^{56}$Fe is plotted as
function of energy of incoming electron $E_{\mathrm{e}^-}$. The
width of a band of obtained values is due to the uncertainty,
associated with fixing the particle-particle parameter
$g_\mathrm{pp}$. We found that the contribution from the imaginary
part of ${M}^{(\mathrm{e}\beta^{+})}$ is small, but it increases
with $E_{\mathrm{e}^-}$, as does the whole modulus of the
$(\mathrm{e}^-,\mathrm{e}^+)$ nuclear matrix element. For the
quantitative analysis of the $(\mathrm{e}^-,\mathrm{e}^+)$ capture
rate, we will consider $E_{\mathrm{e}^-}$ from the range (6.33,\,
50)~MeV
\begin{equation}
|{M}^{(\rm{e}\beta^{+})}| \approx 3\,.  \label{ME}
\end{equation}
%

\section{Energy production of the reaction $\rm{e}^-$\,$\rightarrow$\,$\rm{e}^+$
in the strongly magnetized iron white dwarfs} \label{EPIWD}

To estimate the energy production $\bar{\varepsilon}_r$ per one
event of the reaction (\ref{EMEPC}), we calculated the two-photon
positron-electron annihilation probability per volume within the
quantum electrodynamics framework \cite{W,BD} and integrated it over
the energies of electrons $\epsilon_f=E_f/m_e$ interacting with the
positron in the final state of the reaction (\ref{EMEPC}) according
to the prescription (\ref{SOES}). As a result, we obtained
\be
\bar{\Gamma_i}\,=\,\frac{2\gamma}{(2\pi)^2\lambda^3_\mathrm{e}}\,
\int\limits_{1}^{\epsilon_{\,\mathrm{F}}}\,
\frac{\epsilon_f}{(\epsilon^2_f-1)^{1/2}}\,\Gamma_i\,d\epsilon_f\,,\quad
i=0,1\ .  \label{GB} \ee
Our notations used in this section follow closely those of Chapter~8
in Ref.\,\cite{W}. Further, \be \Gamma_i=
\frac{(2\pi)^4}{8}\,\sum_{\epsilon,\epsilon',\,\sigma,\sigma'}\,\int\,|M|^2\,
\delta^{(4)}(p_f+p_{\mathrm{e}^+}-k-k')f_i(k^0+k'^0)\,d\bm{k}\,d\bm{k}'\,,
\label{G} \ee $f_0(k^0+k'^0)=1$, $f_1(k^0+k'^0)=k^0+k'^0$, the sum
is performed over two photon linear polarizations $\epsilon$,
$\epsilon'$ and the average is done over the electron (positron)
spin z-component $\sigma$ ($\sigma'$), the final photons have the
4-momenta $k^\mu$ and $k'^\mu$ ($k^0=|\bm{k}|=\omega$,
$k'^0=|\bm{k}'|=\omega'$), the positron (electron) 4-momentum is
$p^\mu_{\rm{e}^+}$ ($p^\mu_f$), also $E_{\rm{e}^+}=p^0_{\rm{e}^+}$
and $E_f=p^0_f$. In its turn, the amplitude $M$ is
\bea M\,&=&\,-\frac{e^2}{4(2\pi)^3\,\sqrt(\omega\,\omega')}\, {\bar
v}(\bm{p}_{\rm{e}^+},\sigma')\left\{\slashed{\epsilon}'
[-i(\slashed{p}_f - \slashed{k} )+m_\mathrm{e}]\slashed{\epsilon}\,/(p_f\cdot k) \right.  \nonumber \\
 && \qquad \qquad \qquad \qquad
 \left. +\slashed{\epsilon} [-i(\not p_f - \slashed{k}')+m_\mathrm{e}]\slashed{\epsilon}'
 \,/(p_f\cdot k')\right\}u(\bm{p}_f,\sigma)\,. \label{AM}
\eea
The calculation of the $|M|^2$ reduces to evaluation of traces and
yields
\bea \Gamma_i\,&=&\, \frac{\alpha^2 \pi}{2m^2_\mathrm{e}
\epsilon_{\mathrm{e}^+}}\,\int\limits_{-1}^{+1}\,dx
\,\frac{\bar\omega}{\epsilon_f
+\epsilon_{\mathrm{e}^+}+(|\bar{\bm{p}}_{\mathrm{e}^+}|-|\bar{\bm{p}}_f|)
x}\,\left\{\,A_0+A_1/ (\bar p_f \cdot \bar k)^2
 \right. \nonumber \\
&& \qquad \qquad \qquad  \left.\, + A_2/(({\bar p}_f \cdot {\bar k})
 ({\bar p}_f \cdot {\bar k'}))
+A_3/({\bar p}_f \cdot {\bar k'})^2\,\right\}\,m^i_\mathrm{e}
f_i(\bar\omega+\bar\omega')\ .   \label{GG} \eea
Here, the bared quantities are expressed in units of the electron
mass and
\be \omega\,=\,\frac{m^2_\mathrm{e}+E_{\mathrm{e}^+}
E_f-|\bm{p}_{\mathrm{e}^+}| |\bm{p}_f|}{E_{\mathrm{e}^+}+ E_f
+(|\bm{p}_{\mathrm{e}^+}|- |\bm{p}_f|) x}\,,\quad
\omega'\,=\,E_{\mathrm{e}^+}+ E_f-\omega\ .   \label{OM} \ee
The scalar function $A_0$ comes from those parts of the traces that
do not contain the factors ($\epsilon \cdot p_f$) and ($\epsilon'
\cdot p_f$). The functions $A_i$, $i=0,1,2,3$, are presented in
Appendix \ref{appA}.

In the next step, we include $\bar\Gamma_i$ into  the function
$\phi(\epsilon_{\,\mathrm{F}},\gamma)$ , obtaining thus
$\phi_0(\epsilon_{\,\mathrm{F}},\gamma)$  and
$\phi_1(\epsilon_{\,\mathrm{F}},\gamma)$ . The energy produced in
one event per one second is then calculated as
\be
\bar{\varepsilon}_r\,=\,\phi_1(\epsilon_{\,\mathrm{F}},\gamma)
/\phi_0(\epsilon_{\,\mathrm{F}},\gamma)\ . \label{EPOE}
\ee
Since the Fermi energy $E_{\,\mathrm{F}}$ of the electron gas is
larger than the threshold energy  given by the mass difference between
the final and the initial nuclei $\Delta^{\beta\beta}_{56}$ plus
the positron mass, in average a heat energy in one electron capture
event
\be
\Delta h = (E_{\,\mathrm{F}}-\Delta^{\beta\beta}_{56}-m_\mathrm{e})/2
\ee
is released, which should be added to $\bar{\varepsilon}_r$.

In the interval of 1 year, the number of reactions in 1 cm$^3$ is
$n_r=n_\mathrm{m}\,\Gamma$, where $n_\mathrm{m}$ is the number
density of matter. Then the released energy in 1 cm$^3$ per 1 year
is
\be E=n_\mathrm{m}\,\Gamma\,\bar{\varepsilon}_r\
[\mathrm{J}\,\mathrm{cm}^{-3}\,\mathrm{y}^{-1}]\ . \label{EPOY} \ee
Let us consider the SMIWD with the mass
$M_{\mbox{\tiny{SMWD}}}$=2\,$M_\odot$\,$\approx$\,4\,$\times\,$10$^{33}$\,g.
Its volume is
$V_{\mbox{\tiny{SMWD}}}$=$M_{\mbox{\tiny{SMWD}}}$/$\rho_\mathrm{m}$,
from which one obtains the radius $R_{\mbox{\tiny{SMWD}}}$. In the
full volume, the released energy per 1 year is $EV$ [J y$^{-1}$],
from which one obtains directly the change in the luminosity
(released energy per 1 second) $\Delta L\,=\, EV/3.154\,\times 10^7$
W. The influence on the surface temperature of the white dwarf can
be calculated from the equation \be \Delta T\,=\,\left(\frac{\Delta
L}{s\, 4\pi R_{\mbox{\tiny{SMWD}}}^2}\right)^{1/4}\,. \label{DT} \ee
Here, $s=5.67\,\times\, 10^{-8}$ W\,m$^{-2}$\,K$^{-4}$.

The calculated change in the luminosity and in the surface
temperature of the SMIWDs is presented in Table \ref{tab:res1},
obtained by using the necessary input from Table \ref{tab:inpt},
$|\langle m_\nu\rangle|$=0.4 eV and
Eqs.\,(\ref{vfi}),\,(\ref{RRPNF}),\,(\ref{ME}). Besides, $R=1.2
A^{1/3}\,\approx\,$ 4.59 fm.

\begin{table}[tb]
\caption{The values of the Fermi energy
$\epsilon_{\,\mathrm{F}}=E_{\,\mathrm{F}}/m_\mathrm{e}$ used in the
present study. Further, $R_{\mbox{\tiny{WD}}}$ is the radius of the
SMIWD,  $\Delta\,L$  is the calculated change in the luminosity,
$\Delta\,T$ is the related change in the surface temperature, and
the released energy in single reaction, $\bar{\varepsilon}_r$ is
defined in Eq.\,(\ref{EPOE}); the surface temperature of the SMIWD
$T_{eff}$ is obtained from Eq.\,(\ref{DT}), by taking the luminosity
$L = 10^{-8}\, L_\odot$. }
\begin{center}
\begin{tabular}{|l |  c    | c    |  c    |  c   |  c  | c |}\hline
  $\epsilon_{\,\mathrm{F}}$ & $R_{\mbox{\tiny{SMWD}}}$/10$^4$ [m] & $\Delta\,L$ [W]&
  $\Delta\,T$ [K]   & $T_{eff}$ [K]   & $\bar{\varepsilon}_r$ [MeV] \\\hline
 20        & 42.3 & 5.9 x 10$^{11}$   & 46  &  2340   & 15.6   \\
 46        & 18.6 & 7.4 x 10$^{14}$   & 420 &  3550   & 44.7  \\\hline
\end{tabular}
\end{center}
\label{tab:res1}
\end{table}

Since neither the luminosity, nor the surface temperature of the
SMIWDs are known so far, we estimated possible effect of the
reaction (\ref{EMEPC}), given by $\Delta\,T$ in Table
\ref{tab:res1}, on the surface temperature $T_{eff}$, by taking in
Eq.\,(\ref{DT}) the luminosity $L = 10^{-8}\,L_\odot$, where the
solar luminosity $L_\odot$= 3.828\,$\times$\,10$^{26}$ W \cite{PDG}.
These estimates indicate that the change in the temperature can be
as large as 10  \% of $T_{eff}$. However, as is seen from the values
of $T_{eff}$, such objects are too dim to be observed.

In Table \ref{tab:res2}, we present  the ratio of the calculated
change in the luminosity $\Delta\, L$ of the SMIWDs to the solar
luminosity $L_\odot$, employing again the necessary input from Table
\ref{tab:inpt}, $|\langle m_\nu\rangle|$=0.4 eV and 0.8 eV, and
$R=1.2 A^{1/3}\,\approx\,$ 4.59 fm.

\begin{table}[tb]
\caption{The values of the Fermi energy
$\epsilon_{\,\mathrm{F}}$=$E_{\,\mathrm{F}}/m_\mathrm{e}$ used in
the present study. Further,  $(\Delta L/L_\odot)_{0.4}$ and $(\Delta
L/L _\odot)_{0.8}$ are the ratios of the change in the luminosity of
the SMIWD due to the reaction (\ref{EMEPC}) to the luminosity of the
Sun, calculated for $|\langle m_\nu\rangle|$=0.4 eV and 0.8 eV,
respectively.}
\begin{center}
\begin{tabular}{|l |  c    | c    |   }\hline
  $\epsilon_{\,\mathrm{F}}$ & $\left[Log(\Delta L/L_\odot)\right]_{0.4}$ &
  $\left[Log(\Delta L/L_\odot)\right]_{0.8}$ \\\hline
 20        & -14.8 & -14.2    \\
 46        & -11.7 & -11.1    \\\hline
\end{tabular}
\end{center}
\label{tab:res2}
\end{table}

In the next section, we discuss the cooling of white dwarfs and
compare our results with the existing cooling models.

\section{Cooling of white dwarfs}
\label{cwd}

The basic model of cooling of the WDs was formulated by Mestel
\cite{ST,LM}. In this model, the thin surface layer is considered as
non-degenerate, whereas the interior of the WDs is taken as fully
degenerate.  The accumulated heat energy of the core is transported
to the surface by diffusion of photons and electrons.

Concerning the cooling of the SMWDs, to our knowledge, no specific
calculations were performed till now. Various processes occurring in
the plasma under the influence of a strong magnetic field were
studied already 40 years ago in Refs.\,\cite{CCh,FC,CC,CCC,CLR} and
recently reviewed in Ref.\,\cite{AYPA}.

In order to estimate qualitatively possible effect of the reaction
(\ref{EMEPC}) on the cooling of the SMIWDs, we suggested that the
correctly calculated cooling process would be similar as for the
iron-core WDs. Extrapolating the data, presented in Fig.\,17
\cite{PAB} for the curve corresponding to $M/M_\odot$=0.6 to smaller
values of the luminosity, we got that
\mbox{$Log(L/L_\odot)\,\approx\,$-5.0 (-7.54)} is achieved after the
cooling time $\tau\,\approx\,$3.48 (3.90) Gyr \footnote{We obtained
similar results also extrapolating the data for the curve
corresponding to $M/M_\odot$=0.8.}. It is clear from Table
\ref{tab:res2} that the effect of the double charge exchange
reaction (\ref{EMEPC}) could influence the luminosity only in the
asymptotic region. Since the inverse beta-decay should be taken into
account in the study of the SMIWDs as well, it follows from
Refs.~\cite{BK1,BS,BK2} that the effect of the reaction
(\ref{EMEPC}) on the asymptotic will be combined with the
non-equilibrium heat from the second part of the reaction
(\ref{R1}).

\section{SMIWDs as SGRs/AXPs}
\label{sgr_axp}

During the last decade the observational astrophysics made
substantial progress in the study of the SGR- and AXP sources of the
radiation \cite{MCG}. These compact objects are standardly
identified with magnetars, that are a class of the NSs
powered by strong magnetic fields up to 10$^{15}$ G. In the last
years, SGRs/AXPs with lower magnetic fields of the order $\sim$
10$^{12}$-10$^{13}$ G and with the rotation period P $\sim$ 10 s
have been observed. It was shown in Refs.\,\cite{MRR,CM,BIRR} that these
fast rotating magnetized NSs can be alternatively
described as massive fast rotating magnetized WDs. Here we show that
similar description is possible within the concept of the fast
rotating SMIWDs. Below we follow closely the notations of Section 2
\cite{CM}.

The magnetic moment of the rotating star can be expressed in terms
of the observables  such as the rotational period $P$ and the
spin-down rate $\stackrel{.}{P}$ as \be m=\left(\frac{3 c^3 I}{8
\pi^2}\,P {\stackrel{.}{P}} \right)^{1/2}\,, \label{MM} \ee where
$I$ is the momentum of inertia. Besides, the surface magnetic field
at the equator is \be B_e = \frac{m}{R^3}\,,  \label{BE} \ee where
$R$ is the radius of the star at the equator.

In this pulsar model, the X-ray luminosity is supposed to come fully
from the loss of the rotational energy \be \stackrel{.}{E}_{rot} =
-4\pi^2 I \frac{\stackrel{.}{P}}{P^3}\,,  \label{ERO} \ee and the
characteristic age of the pulsar is \be \tau =
\frac{P}{2\stackrel{.}{P}}\,.   \label{AGE} \ee In the case of the
magnetar model \cite{DT,TAD}, the choice for the mass of the NS is
$M = 1.4 M_\odot$ and $R$ = 10 km, whereas in the WD model
\cite{MRR,CM,BIRR}, $M = 1.4 M_\odot$ and $R$ = 3000 km.

In the approach of the SMIWDs we take $M=2 M_\odot$ and the radii
from our Table \ref{tab:res1}. Next we analyze the data for the
magnetars \mbox{SGR 0418+579} and \mbox{Swift J1822.6-1606}.

a) \underline{SGR 0418+5729}

\bea
 P(s)&=&9.0784\,\,^a)\,,\quad \stackrel{.}{P}(s\,s^{-1}) = 4\times 10^{-15}\,\,^a)\,,
\quad d=2\, {\rm kpc}\,^a)\,\,,  \nonumber \\
 \Delta L_X&=&7.5\times 10^{-15}\,{\rm erg\, s}^{-1} {\rm cm}^{-2}\,^b)\,\,.\label{DSGR}
\eea
$^a$)\,\,Ref.\,\cite{RSGR}\,,\quad $^b$)\,\,Ref.\,\cite{RPC}  \

From $\Delta L_X$  and the distance $d$ of Eq.\,(\ref{DSGR}) one
obtains for the total luminosity and the age \be L_X = 3.6 \times
10^{30} {\rm erg s}^{-1}\,,\quad \tau = 36\, {\rm Myr}\,.
\label{LXTSG} \ee Using Eqs.\,(\ref{MM})-(\ref{AGE}) and the data
(\ref{DSGR}) one arrives at the results \be m_{\tiny{NS}} =
6.4\times 10^{30}\,{\rm emu}\,,\quad B_{\tiny{NS}} = 6.4\times
10^{12}\,{\rm G}\,,\quad |\stackrel{.}{E}_{\tiny{rot}}^{\tiny{NS}}|
= 7.5\times 10^{28} {\rm erg\, s}^{-1}\,,   \label{SGRNS} \ee \be
m_{\tiny{WD}} = 1.9\times 10^{33}\,{\rm emu}\,,\quad B_{\tiny{WD}} =
7.1\times 10^{7}\,{\rm G}\,,\quad
|\stackrel{.}{E}_{\tiny{rot}}^{\tiny{WD}}| = 6.7\times 10^{33} {\rm
erg\, s}^{-1}\,.   \label{SGRWD} \ee
\begin{table}[tb]
\caption{The values of the magnetic moment $m_{\tiny{SMIWD}}$ (in
emu), of the flux of the magnetic field $B_{\tiny{SMIWD}}$ (in G)
and of the rotational energy
$|\stackrel{.}{E}_{\tiny{rot}}^{\tiny{SMIWD}}|$ (in erg s$^{-1}$)
for the compact object SGR 0418+5729, considered as a fast rotating
SMIWD.}
\begin{center}
\begin{tabular}{|l |  c    | c     |}\hline
  $R_{\mbox{{\tiny{SMIWD}}}}$ [km]& 423 & 186  \\\hline
 $m_{\mbox{{\tiny{SMIWD}}}}/10^{31}$ & 32.5 & 14.3    \\
 $B_{\mbox{{\tiny{SMIWD}}}}/10^{10}$ & 0.43 & 2.22    \\
 $|\stackrel{.}{E}_{\tiny{rot}}^{\tiny{SMIWD}}|/10^{30}$ & 191  & 37  \\\hline
\end{tabular}
\end{center}
\label{tab:res3}
\end{table}
Comparing the results for the rotational energy, presented in Table
\ref{tab:res3}, with the total luminosity (\ref{LXTSG}) one can see
that the loss of this energy of the SMIWDs can explain $L_X$. As is
seen from Eq.\,(\ref{SGRWD}), this is also true for the fast
rotating WD. On the contrary, the loss of the rotational energy of
the NS (\ref{SGRNS}) is by about two orders of the magnitude smaller
than~(\ref{LXTSG}).

\newpage

b) \underline{Swift J1822.6-1606}

The data below are taken from Ref.\,\cite{RSWIFT}:

\bea
 P(s)&=&8.4377\,\,,\quad \stackrel{.}{P}(s\,s^{-1}) = 8.3\times 10^{-14}\,\,,
\quad d=5\, {\rm kpc}\,,  \nonumber \\
 \Delta L_X&=&4\times 10^{-14}\,{\rm erg\, s}^{-1} {\rm cm}^{-2}\,.\label{DSWIFT}
\eea From $\Delta L_X$  and the distance $d$ of Eq.\,(\ref{DSWIFT})
one obtains for the total luminosity and the age \be L_X = 1.2
\times 10^{32} {\rm erg s}^{-1}\,,\quad \tau = 1.61\, {\rm Myr}\,.
\label{LXTSW} \ee Using Eqs.\,(\ref{MM})-(\ref{AGE}) and the data
(\ref{DSWIFT}) one arrives at the results \be m_{\tiny{NS}} =
2.8\times 10^{31}\,{\rm emu}\,,\quad B_{\tiny{NS}} = 2.8\times
10^{13}\,{\rm G}\,,\quad |\stackrel{.}{E}_{\tiny{rot}}^{\tiny{NS}}|
= 1.9\times 10^{30} {\rm erg\, s}^{-1}\,,   \label{SWIFTNS} \ee \be
m_{\tiny{WD}} = 8.5\times 10^{33}\,{\rm emu}\,,\quad B_{\tiny{WD}} =
3.1\times 10^{8}\,{\rm G}\,,\quad
|\stackrel{.}{E}_{\tiny{rot}}^{\tiny{WD}}| = 1.7\times 10^{35} {\rm
erg\, s}^{-1}\,.   \label{SWIFTWD} \ee
\begin{table}[tb]
\caption{The values of the magnetic moment $m_{\tiny{SMIWD}}$ (in
emu), of the flux of the magnetic field $B_{\tiny{SMIWD}}$ (in G)
and of the rotational energy
$|\stackrel{.}{E}_{\tiny{rot}}^{\tiny{SMIWD}}|$ (in erg s$^{-1}$)
for the compact object Swift J1822.6-1606, considered as a fast
rotating SMIWD.}
\begin{center}
\begin{tabular}{|l |  c    | c     |}\hline
 $R_{\mbox{{\tiny{SMIWD}}}}$ [km]& 423 & 186   \\\hline
 $m_{\mbox{{\tiny{SMIWD}}}}/10^{32}$ & 14.3 & 6.28   \\
 $B_{\mbox{{\tiny{SMIWD}}}}/10^{11}$ & 0.19 & 0.98    \\
 $|\stackrel{.}{E}_{\tiny{rot}}^{\tiny{SMIWD}}|/10^{32}$ & 49.4  & 9.6  \\\hline
\end{tabular}
\end{center}
\label{tab:res4}
\end{table}

Comparing the results for the rotational energy, presented in Table
\ref{tab:res4}, with the total luminosity (\ref{LXTSW}) one can see,
as in the case of SGR 0418+5729, that the loss of this energy of the
SMIWDs can explain $L_X$. As is seen from Eq.\,(\ref{SWIFTWD}), this
is also true for the fast rotating WD. On the contrary, the loss of
the rotational energy of the NS (\ref{SWIFTNS}) is by about two
orders of the magnitude smaller than (\ref{LXTSG}).

On the other hand, comparison of the rotational energies of
Table\,\ref{tab:res3} and Table\,\ref{tab:res4} with the 3rd column
of Table\,\ref{tab:res1} shows that the energy produced by the
reaction of the double charge exchange (\ref{IBD}) cannot influence
sizeably the luminosity of the compact objects considered above as
fast rotating SMIWDs.


\section{Discussion of the results and conclusions}
\label{concl}

In this work, we studied the influence of the double charge exchange
reaction (\ref{EMEPC}) on the cooling of the SMIWDs. This reaction
is closely related to the neutrinoless double beta-decay process,
which is nowadays studied intensively \cite{VES}. Both processes
violate the lepton number by two units and, therefore, take place if
and only if the neutrinos are the Majorana particles with the
non-zero mass. For the case of light neutrino exchange, the
$(\mathrm{e}^-,\mathrm{e}^+)$ conversion and the
$0\nu\beta\beta$-decay rates are proportional to the squared
absolute value of the effective mass of the Majorana neutrinos,
$|\langle m_\nu\rangle|^2$.

Our study is based on the theory of the SMWDs \cite{KM,DM,DM1,DM2}.
This theory, when applied to the phenomenon of the Ia supernovae,
can reasonably explain the existence of the observed progenitor star
with the mass exceeding the CL limit \cite{DM3,DM4}. Our model
calculations are done for the case of the SMIWDs, in which the
magnetic field is strong enough to maintain the Fermi energy of the
electron sea larger than the threshold energy for the reaction
(\ref{EMEPC}), which is 6.33 MeV. We considered
$\epsilon_{\,\mathrm{F}}=E_{\,\mathrm{F}}/m_\mathrm{e}$=20,\,46, and
have chosen the strength of the magnetic field so that the value of
the related parameter $\gamma={\cal B}/{\cal B}_c$ allowed us to
restrict ourselves to the ground Landau level.

The calculations are first performed under the assumption that the
SMIWDs radiate in the visible spectrum. The results are presented in
Table \ref{tab:res1} and Table \ref{tab:res2}.

In calculating Table \ref{tab:res1}, we suggested that an SMIWD can
possess the luminosity of the size $L = 10^{-8} L_\odot$ and
obtained the corresponding surface temperature from Eq.\,(\ref{DT}).
Then the comparison with the calculated effect of the reaction
(\ref{EMEPC}) shows possible influence of 10 \%. However, as is seen
from the values of $T_{eff}$, such objects would be too dim to be
observed.

Since the theory of cooling of the SMIWDs has not yet been
developed, we turned to the one for the iron-core WDs, which is well
elaborated. By comparing the luminosity of the pure iron-core DA
models of Fig.\,17 \cite{PAB} with our Table \ref{tab:res2} we can
conclude that the double charge exchange reaction (\ref{EMEPC})
could in the case of the SMIWDs retard their cooling at low
luminosity by pumping over the energy of the Fermi sea of electrons
to the thermal energy of ions. However, the effect would be out of
reach of the present observational possibilities. Moreover, it would
be diminished by the non-equilibrium heat from the inverse
beta-decay reaction (\ref{R1}) \cite{BK1,BS,BK2}.

Then in Section \ref{sgr_axp}, we explored  the SMIWDs as fast
rotating stars that can be considered as GSR/AXPs. We have shown (
see \mbox{ Table \ref{tab:res3}} and \mbox{Table \ref{tab:res4}})
that using the observational data for the magnetars \mbox{SGR
0418+579} and \mbox{Swift J1822.6-1606}, the calculated loss of the
rotational energy can reproduce the observed total luminosity for
the considered SWIMDs. However, the energy produced by the reaction
of the double charge exchange (\ref{IBD}) cannot influence sizeably
the luminosity of the compact objects considered as fast rotating
SMIWDs.

Our main conclusion is that the energy, released in reaction
(\ref{EMEPC}), (i) could influence under certain assumption the
effective temperature up to 10 \%, but the SWIMDs would be too dim
to be observed at present; (ii) could retard cooling of the SMIWDs
at sufficiently low luminosity, which seem to be, however, at an
unobservable level at present as well; (iii) cannot influence
sizeably the luminosity of these compact objects considered as the
sources of SGR/AXP radiation. It means that the study of the double
charge exchange reaction (\ref{EMEPC}) in the SMIWDs using simple
model \cite{KM,DM,DM1,DM3,DMR} with the ground Landau level and at
the present level of accuracy of measurement of the luminosity and
energy of the cosmic gamma-rays could not provide conclusive
information on the Majorana nature of the neutrino, if its effective
mass would be $|\langle m_\nu \rangle|$ $\le$ 0.8 $eV$.

\section*{Acknowledgments}
The work of V.~Belyaev was supported  by the Votruba-Blokhintsev
Program for Theoretical Physics of the Committee for Cooperation of
the Czech Republic with JINR, Dubna.  F. \v Simkovic acknowledges
the support by the VEGA Grant agency of the Slovak Republic under
the contract No. 1/0876/12. The research of M. Tater was supported
by the Czech Science Foundation within the project P203/11/0701. We
thank B.~Mukhopadhyay for correspondence, L.~Althaus for
providing us with the data, presented in Fig.\,17 \cite{PAB} by the
luminosity curves for the pure iron-core DA WDs and N.~Rea for
communicating us the data on \mbox{SGR 0418+579}.




\bibliographystyle{aipproc}   

\bibliography{bibliography}

\IfFileExists{\jobname.bbl}{}
 {\typeout{}
  \typeout{******************************************}
  \typeout{** Please run "bibtex \jobname" to optain}
  \typeout{** the bibliography and then re-run LaTeX}
  \typeout{** twice to fix the references!}
  \typeout{******************************************}
  \typeout{}
 }

\newpage

\appendix

\section{Calculations of traces}
\label{appA}

Here we provide the invariant functions $A_i$ entering the
positron-electron annihilation probability (\ref{GG}), resulting
from calculations of traces and summing over the photon linear
polarizations. The calculations are made in the Coulomb gauge,
putting $\epsilon^0$=$\epsilon'^0$=0 and using \be
\sum_{\epsilon}\epsilon_i \epsilon_j\,=\,\delta_{ij}-\vk_i \vk_j\,,
\quad \sum_{\epsilon'}\epsilon'_i\,\epsilon'_j\,=\,\delta_{ij}
-\vkp_i \vkp_j\,.  \nonumber \\
\ee

\be
m^4_\mathrm{e} A_0\,=\,\frac{(k\cdot k')^2}{(p_f\cdot k)(p_f\cdot k')}
- a_0\,,  \label{A0}
\ee
\be
m^4_\mathrm{e} A_1\,=\,\left\{(-a_3+a_5+a_{11})(p_f\cdot k)+[\,(p_f\cdot k')
-(k\cdot k')+a_1-a_2\,] a_{13}\right\}
/(p\cdot k)^2\,,  \label{A1}
\ee
\bea
m^4_\mathrm{e} A_2\,&=&\,\left\{[a_1-a_4+a_{12}-a_{13}+(a_8-a_6)/2](p_f\cdot k)
+\left[a_3-a_{10}-a_{11}+a_{12} \right. \right. \nonumber \\
 &&  \left.\left.  \,+(a_6-a_8)/2\right](p_f\cdot k') +[-a_1-a_3+a_4
 +a_{11}-a_{12}+(a_5+a_7)/2](k\cdot k')  \right. \nonumber \\
 &&  \left. \, +(a_9-a_3)a_{11}+(a_{13}-a_4)a_1+a_3 a_{10}\right\}
 /(p_f\cdot k)(p_f\cdot k')\,,   \label{A2}
\eea \be m^4_\mathrm{e} A_3\,=\,\left\{[\,(p_f\cdot k)-(k\cdot
k')+a_3-a_9\,]a_{10} +(-a_1+a_4+a_7)(p_f\cdot k')\right\}/(p_f\cdot
k')^2\,.   \label{A3} \ee Further, \be a_0\,=\,1 +
(\vk\cdot\vkp)^2\,,   \label{a0} \ee \be
a_1\,=\,(\vp\cdot\vpp)-(\vp\cdot\vkp)(\vpp\cdot\vkp)\,,  \label{a1}
\ee \be a_2\,=\,(\vpp\cdot\vk)-(\vpp\cdot\vkp)(\vk\cdot\vkp)\,,
\label{a2} \ee \be
a_3\,=\,(\vp\cdot\vpp)-(\vp\cdot\vk)(\vpp\cdot\vk)\,,  \label{a3}
\ee \be a_4\,=\,(\vp\cdot\vkp)-(\vp\cdot\vk)(\vk\cdot\vkp)\,,
\label{a4} \ee \be
a_5\,=\,(\vp\cdot\vpp)-(\vp\cdot\vk)(\vpp\cdot\vk)-(\vp\cdot\vkp)(\vpp\cdot\vkp)
+(\vp\cdot\vk)(\vpp\cdot\vkp)
        (\vk\cdot\vkp)\,,   \label{a5}
\ee
\be
a_6\,=\,(\vk\cdot\vkp)\left[-(\vp\cdot\vkp)+(\vp\cdot\vk)(\vk\cdot\vkp)\right]
\,,   \label{a6}
\ee
\be
a_7\,=\,(\vp\cdot\vpp)-(\vp\cdot\vk)(\vpp\cdot\vkp)-(\vp\cdot\vkp)(\vpp\cdot\vkp)
+(\vp\cdot\vkp)(\vpp\cdot\vk)
        (\vk\cdot\vkp)\,,   \label{a7}
\ee \be
a_8\,=\,(\vk\cdot\vkp)\left[-(\vp\cdot\vk)+(\vp\cdot\vkp)(\vk\cdot\vkp)\right]
\,,   \label{a8} \ee \be
a_9\,=\,(\vpp\cdot\vkp)-(\vpp\cdot\vk)(\vk\cdot\vkp)\,,   \label{a9}
\ee \be a_{10}\,=\,\vp^2-(\vp\cdot\vkp)^2\,,   \label{a10} \ee \be
a_{11}\,=\,(\vp\cdot\bm k)-(\vp\cdot\vkp)(\bm k\cdot\vkp)   \,,
\label{a11} \ee \be
a_{12}\,=\,\vp^2-(\vp\cdot\vk)^2-(\vp\cdot\vkp)^2+(\vp\cdot\vk)(\vp\cdot\vkp)
(\vk\cdot\vkp)\,,  \label{a12} \ee \be
a_{13}\,=\,\vp^2-(\vp\cdot\vk)^2\,.   \label{a13} \ee Here,
${\hat{\bm a}}=\bm a/|\bm a|$ is the unit vector. The invariant
function $A_0$ arises from the part of traces that do not contain
the factors $(\epsilon\cdot p_f)$ and $(\epsilon'\cdot p_f)$. If one
puts $\bm p_f=0$, one obtains the positron-electron annihilation
probability in the laboratory frame of reference. At threshold
positron energies one gets for $\Gamma_0$ \be
\Gamma_0\,=\,\frac{\alpha^2 \pi}{m_\mathrm{e}^2}\,,  \label{G0t} \ee
which provides for the annihilation cross section \be
\sigma\,=\,\frac{\alpha^2 \pi}{v m_\mathrm{e}^2}\,,  \label{si} \ee
where $v$ is the positron velocity.

\end{document}